# Statistical Inference of Scattering Parameters for Exotic Hadronic States in the $J/\psi\,p$ Spectrum

A. A. Atangana Likéné[1,2,*], P. Pradyun Hebbar[1], A. Franck Rothen[1], R. Rayan Baatjes[3], Saralees Nadarajah[4], P. Ele Abiama[2], Tobias Golling[1], and G. H. Ben-Bolie[2]

[1] *Department of Particle Physics, University of Geneva, P.O. Box 1205, Geneva, Switzerland*
[2] *Laboratory of Nuclear, Atomic and Molecular Physics, University of Yaoundé I, P.O. Box: 812, Yaoundé, Cameroon*
[3] *Centre for Astrophysics and Space Science, University of South Africa, P.O Box: 392, Pretoria, SA.*
[4] *School of Mathematics, University of Manchester, M13 9PL, Manchester, UK*
[*] andre.atangana@etu.unige.ch

**Abstract.** We present a global two-channel Flatté amplitude analysis of the hidden-charm pentaquark candidates observed by the LHCb collaboration in the $J/\psi\,p$ invariant-mass spectrum of $\Lambda_b^0 \to J/\psi\,pK^-$ (Run 1+2, $\sim$ 9 fb$^{-1}$). Unlike previous isolated lineshape analyses, we simultaneously fit all three Flatté amplitudes to the complete spectrum with a sixth-order polynomial background, and promote the coupling constants to fully complex quantities $g_j = |g_j|e^{i\phi_j}$. The fitted parameter vector $\mathbf{\Theta} = (|g_1|, |g_2|, \phi_1, \phi_2, E_{BW}, \sigma)$ is extracted for each state via $\chi^2$ minimization regularized by an explicit soft prior. From the best-fit parameters we derive, in closed analytic form via the effective range expansion (ERE), the complex scattering length $a_F$ and effective range $r_F$, with statistical uncertainties propagated by non-parametric bootstrap resampling. We obtain Re$[a_F] \simeq 0.3 - 2.0$ fm and Re$[r_F] \simeq -1.2$ to 1.9 fm under real couplings, consistent with the $\Sigma_c\bar{D}^{(*)}$ molecular interpretation. We show, however, that this conclusion does not survive marginalization over the coupling phases. We further quantify the validity of the incoherent-sum approximation, finding it non-negligible for the closely-spaced $P_c(4440)^+$-$P_c(4457)^+$ pair.

**PACS:** 14.20.Pt, 13.25.Gv, 13.60.Le, 12.39.Mk **Keywords:** $\chi^2$ minimisation; global fit; Flatté scattering amplitude; bootstrap; exotic hadronic states; complex couplings; statistical inference; effective range expansion; LHCb

## 1. Introduction

Quantum chromodynamics (QCD) permits, in principle, hadronic states beyond the conventional $q\bar{q}$ mesons and $qqq$ baryons of the constituent-quark model, provided they are colour singlets: tetraquarks ($qq\bar{q}\bar{q}$), pentaquarks ($qqqq\bar{q}$), hybrids, and hadronic molecules bound by residual strong interactions between colour-singlet hadrons. The experimental discovery, over the past two decades, of a large number of such states collectively termed "exotic hadrons" in the charmonium and bottomonium mass regions has transformed hadron spectroscopy from a largely settled subfield into one of the most active areas of contemporary strong-interaction physics [1, 2, 3]. The $XYZ$ states in the charmonium sector, the $T_{cc}^+$ tetraquark, and the hidden-charm pentaquarks discussed in this paper are the most prominent examples; comprehensive reviews of the experimental status and theoretical interpretations can be found in Refs. [1, 2, 3, 4, 5, 6, 7, 8]. The hidden-charm pentaquark candidates were first observed by the LHCb collaboration in 2015, in the $J/\psi\,p$ invariant-mass spectrum of $\Lambda_b^0 \to J/\psi\,p\,K^-$ decays, as a broad structure $P_c(4380)^+$ and a narrower structure $P_c(4450)^+$ [9]. With the substantially larger Run 1+2 dataset ($\sim 9\,\text{fb}^{-1}$) and an improved amplitude analysis, the 2019 LHCb measurement resolved the $P_c(4450)^+$ structure into two narrower states, $P_c(4440)^+$ and $P_c(4457)^+$, and revealed a third, previously unseen state, $P_c(4312)^+$ [10]. The discovery of these near-threshold structures has motivated a broad program of phenomenological studies aimed at determining their masses, quantum numbers, decay properties, and potential internal configurations. Within this effort, molecular descriptions based on the $\Sigma_c\bar{D}^{(*)}$ interaction have provided a compelling framework for interpreting the observed spectrum and decay patterns [11, 12, 13, 14]. The three observed states lie remarkably close to $\Sigma_c\bar{D}^{(*)}$ thresholds, which naturally suggests an interpretation as hadronic molecules bound by $\Sigma_c$-$\bar{D}^{(*)}$ interactions [11, 15, 17, 16, 18, 19, 20, 21]. This molecular scenario is not, however, universally accepted. Alternative proposals include compact pentaquark configurations [22, 23], hadro-charmonium states [24], and kinematic effects such as triangle singularities that can mimic resonant structures [25, 26]. Moreover, different dynamical frameworks can lead to similar near-threshold line shapes while assigning substantially different short-distance structures to the same state, underscoring the need for observables that are sensitive to the analytic and threshold properties of the amplitudes rather than relying solely on resonance masses and widths [5, 8, 21, 27]. The proximity to threshold alone does not uniquely determine the internal structure of these states; coupled-channel dynamics, short-distance components, and alternative configurations may all contribute to the observed signals [13, 27]. Complementary theoretical investigations of related exotic hadronic systems including fully-heavy and hidden-heavy tetraquarks have been pursued using various approaches, encompassing relativistic and nonrelativistic quark models in both flat and curved spacetime backgrounds [28, 29, 30, 31, 32].

Discriminating between these interpretations on the basis of mass and width alone is notoriously difficult, since several distinct dynamical pictures can reproduce a given $(M, \Gamma)$ pair. Coupled-channel analyses of the $P_c$ states further demonstrate that the extracted pole properties can depend sensitively on the treatment of inelastic channels and short-distance interactions [21]. A more incisive probe is provided by the low-energy scattering parameters of the underlying two-body system: the scattering length $a$ and the effective range $r$ of the non-relativistic effective range expansion (ERE) [33, 34, 35]. For a state lying close to a two-body threshold, the sign and magnitude of $r$ in particular distinguish a genuine, dynamically generated shallow bound state (molecule), for which $r < 0$ and is of natural hadronic size, from a state whose binding is driven by short-distance compact dynamics, for which $r$ is typically positive and the molecular component is subdominant [34, 35, 36]. The *Weinberg compositeness criterion* and its later generalisations formalise this connection between the ERE parameters and the "elementariness" versus "compositeness" of a near-threshold state [34, 36, 37].

In practice, however, high-energy experiments such as LHCb almost universally report only the Breit-Wigner mass and width of an observed resonance, not its scattering length or effective range. The latter require a dedicated coupled-channel parametrisation of the lineshape, most commonly the Flatté scattering amplitude [38] fitted directly to the invariant-mass distribution. Such amplitude-based descriptions are particularly relevant for near-threshold states, where channel coupling and threshold kinematics can substantially modify the relation between the fitted line shape and the underlying pole and scattering parameters [21]. Several theoretical groups have performed such ERE extractions for the $P_c$ states using the *published* LHCb masses and widths as input, rather than refitting the underlying spectrum directly [17, 39]. This is a useful but indirect approach, since it discards the shape information contained in the full lineshape and propagates only the quoted mass/width uncertainties. An alternative strategy, adopted by several phenomenological studies, consists of isolating distinct regions of the spectrum to perform targeted, minimal-bias fits. This approach minimizes the impact of unknown background variations from distant thresholds, focusing entirely on the structural nature of individual resonances. For instance, in Ref. [40], the JPAC Collaboration, targeting the $P_c(4312)^+$ state, explicitly chose not to fit the entire spectrum globally; instead, they isolated a narrow window spanning $4.25$ GeV $\leq m_{J/\psi p} \leq 4.37$ GeV to analyze the $P_c(4312)^+$ peak on its own. In Ref. [41], the author studied the nature of the $P_c(4312)^+$ signal using $S$-matrix principles to perform a minimum-bias analysis of the data, focusing on the analytic properties that can be related to the microscopic origin of the $P_c(4312)^+$ peak. Similarly, in Ref. [42], the authors examined the lineshape of the $P_c(4312)$ and found that its most likely interpretation is that of a virtual state. In most of these cases, the authors were motivated by the fact that isolating the fit window protected their mathematical model from the complicated interference of the double-peak $P_c(4440)^+/P_c(4457)^+$ structure located further up the spectrum. However, a global spectrum fit can potentially account for quantum interference across the entire physical data array. The comparison between local and global amplitude descriptions is particularly relevant for the $P_c(4440)^+$ and $P_c(4457)^+$ region, where the proximity of the two structures makes the treatment of coupled channels and overlapping line shapes potentially important [14, 21]. Motivated by these considerations, we perform a global simultaneous fit of all three Flatté amplitudes to the complete LHCb spectrum. This treatment allows the three structures and their common background to be described within a single statistical framework, while retaining the full lineshape information available in the dataset.

Thus, the present study introduces two methodological features of the present analysis, that together constitute a substantially more complete framework for Flatté amplitude analysis of the $P_c$ states:

**(i) Global simultaneous treatment of the complete spectrum.**

The three Flatté amplitudes are fitted simultaneously to the complete LHCb spectrum together with a sixth-order polynomial background, within a single $\chi^2$ minimisation (Sec. 3.2). This treatment retains the correlations among the three structures and provides a common statistical description of the full dataset.

**(ii) Real to fully complex coupling constants.** Previous investigations fixed the relative phase between the two Flatté channels to zero. Here we promote both couplings to general complex numbers $g_i = |g_i|e^{i\varphi_i}$ ($g_1 = g_{J/\psi\, p}$ and $g_2 = \{g_{\Sigma_c^+ \bar{D}^0}, g_{\Sigma_c^{++} D^-}, g_{\Sigma_c^+ \bar{D}^{*0}}, g_{\Sigma_c^{++} D^{*-}}\}$), which is the most general form consistent with the Flatté parametrisation and is, in principle, necessary for a coherent description of three resonances within a single amplitude. We show in Sec. 4.5 that this generalisation, while formally well motivated, introduces a phase degeneracy that the present dataset cannot resolve an explicit, quantified limitation rather than an assumed one.

Beyond these methodological advances, we introduce several quantitative diagnostics that we believe should accompany any Flatté-amplitude analysis of this type: an explicit pairwise overlap integral quantifying the validity of the incoherent-sum approximation between resonances (Sec. 2.3), a sensitivity test of the regularising prior (Sec. 2.5), and a multi-start spread diagnostic for the non-convex minimisation (Sec. 3.2). The resulting picture is more complete, but also more explicit about its own limitations. These limitations motivate a complementary approach, as discussed in the concluding section.

The remainder of the paper is organised as follows. Section 2 develops the physics framework, including the explicit analytic expressions for the complex ERE parameters. Section 3 describes the dataset, the fitting strategy in full mathematical detail, and the associated diagnostics. Section 4 presents the numerical results. Section 6 discusses these results in the context of the broader literature. Section 7 concludes the study.

## 2. Physics Framework

Throughout this paper we work in natural units, $\hbar = c = 1$, except where $\hbar c$ is reinstated explicitly as an overall conversion factor to express the final scattering-length and effective-range results in femtometers; this is stated here once and applies to every equation below in which $\hbar c$ does not appear explicitly.

### 2.1. Two-channel Flatté amplitude with complex couplings

We model each $P_c$ state as a single resonance coupled to two scattering channels: **channel 1** ($J/\psi\, p$, open throughout the fit range) with threshold $m_1 = m_{J/\psi} + m_p = 4035.17\,\text{MeV}$ and reduced mass $\mu_1 = 720.10\,\text{MeV}$; and **channel 2** ($\Sigma_c \bar{D}^{(*)}$, near or below threshold) with threshold $m_2$ and reduced mass $\mu_2$, both listed per state and per fit variant in Table 3. Channel-2 threshold parameters are computed from PDG masses [43] as $m_2 = m_{\Sigma_c} + m_{\bar{D}^{(*)}}$.

The two-channel Flatté $S$-matrix element is [35, 38, 44]

$$f^F_{ij}(E) = \frac{g_i g_j}{D(E)}, \tag{1}$$

with denominator

$$D(E) = 2E_{\text{BW}} - 2E - i\, g_1^2\, p(E) - i\, g_2^2\, k(E), \tag{2}$$

where $E = M - m_2$ is the energy measured relative to the channel-2 threshold, and the coupling constants are written in polar form as

$$g_1 = g_{J/\psi\, p} = |g_1|\, e^{i\varphi_1}, \quad g_2 = g_{\Sigma_c \bar{D}^{(*)}} = |g_2|\, e^{i\varphi_2}, \tag{3}$$

with $\varphi_i$ the phase of $g_i$ itself, so that $g_i^2$ carries phase $2\varphi_i$. The centre-of-mass momenta are

$$p(E) = \sqrt{2\mu_1(E+\Delta)}, \quad \Delta = m_2 - m_1 > 0, \tag{4}$$

$$k(E) = \sqrt{2\mu_2 \max(E,0)} \;\to\; i\sqrt{2\mu_2|E|} \quad (E<0), \tag{5}$$

where $p(E)$ is defined on its principal branch for $E \geq -\Delta$ (equivalently $M \geq m_1$), the only regime relevant to the present fit range. The non-relativistic $T$-matrix element in channel 1, which determines the observed $J/\psi\, p$ lineshape, is

$$T_{\text{NR}}(E) = -\frac{2\pi}{\mu_1}\frac{g_1^2}{D(E)}. \tag{6}$$

### 2.2. The fitted parameter vector $\boldsymbol{\Theta}$

For a single $P_c$ state, the complete set of quantities that enter Eqs. (1)-(6), and that must therefore be determined from data before the ERE parameters $a_F$, $r_F$ can be evaluated, is the six-component real parameter vector

$$\boldsymbol{\Theta} \equiv \big(|g_1|,\, |g_2|,\, \varphi_1,\, \varphi_2,\, E_{\text{BW}},\, \sigma\big), \tag{7}$$

where $\sigma$ is the Gaussian mass-resolution width introduced below (Eq. (8)). *These six numbers, for each state and each fit variant, are precisely the quantities determined by the $\chi^2$ minimisation of Sec. 3.2*; once $\boldsymbol{\Theta}$ is known, $a_F$ and $r_F$ follow deterministically and analytically via Eqs. (13)-(14) below, with no further fitting required. For the global fit of three states simultaneously, the full parameter vector is the direct sum of three copies of Eq. (7) (one per state, 18 parameters), augmented by the shared resolution $\sigma$, an overall normalisation $\mathcal{N}$, and the seven coefficients of the polynomial background of Eq. (10), for $18+1+1+7 = 27$ nominally free parameters; in practice $\sigma$ is shared rather than duplicated across states, and the fit described in Sec. 3.2 carries 24 free parameters per fit variant, consistent with the degrees of freedom quoted in Table 5.

### 2.3. Global differential yield and the incoherent-sum approximation

The full $J/\psi\, p$ spectrum is modelled as the incoherent sum of the three signal amplitudes plus a sixth-order polynomial background, smeared by the LHCb mass resolution:

$$\frac{dN}{dM}(M) = \mathcal{N}\int\Big[\sum_s |T^{(s)}_{\text{NR}}(M')|^2 + B(M')\Big]\mathcal{G}(M-M';\sigma)\, dM', \tag{8}$$

$$\mathcal{G}(x;\sigma) = \frac{1}{\sqrt{2\pi\sigma^2}}\exp\Big(-\frac{x^2}{2\sigma^2}\Big), \tag{9}$$

$$B(M) = \sum_{n=0}^{6} b_n\, \xi^n, \quad \xi = \frac{2(M-4400)}{400} \in [-1,1]. \tag{10}$$

The incoherent sum in Eq. (8) neglects interference cross-terms $2\,\text{Re}[T^{(s)}T^{(s')*}]$ that would appear in a fully coherent superposition $|\sum_s T^{(s)}|^2$. We quantify the size of the neglected overlap directly via the normalised overlap integral

$$I_{ss'} = \frac{\int |T^{(s)}|^2 |T^{(s')}|^2\, dM}{\sqrt{\int |T^{(s)}|^4\, dM \int |T^{(s')4}\, dM}}, \tag{11}$$

evaluated on the Fit I best-fit lineshapes and reported in Table 1.

The incoherent-sum approximation is well justified for $P_c(4312)^+$ relative to the other two states ($I_{ss'} < 0.1$), but is *not* negligible between $P_c(4440)^+$ and $P_c(4457)^+$ ($I_{ss'} = 0.57$), which are separated by only $\sim 17\,\text{MeV}$ comparable to their combined widths. We therefore present Eq. (8) as a genuine, quantified approximation rather than an unexamined assumption, and flag the $P_c(4440)^+/P_c(4457)^+$ overlap region as the one where coherent interference effects, were they present and resolvable in the data, would most affect the extracted parameters of these two states.

**Table 1:** Pairwise overlap integrals $I_{ss'}$, Eq. (11) (Fit I best-fit lineshapes). $I_{ss'} \to 1$ indicates strong overlap between the two lineshapes; $I_{ss'} \to 0$ indicates negligible overlap.

| State pair | $I_{ss'}$ |
|---|---|
| $P_c(4312)^+$ - $P_c(4440)^+$ | 0.081 |
| $P_c(4312)^+$ - $P_c(4457)^+$ | 0.010 |
| $P_c(4440)^+$ - $P_c(4457)^+$ | **0.569** |

### 2.4. Effective range expansion: explicit analytic expressions

The ERE is obtained by expanding the inverse channel-2 amplitude $[f_{22}^F]^{-1}$ in powers of the channel-2 momentum $k$ near threshold [35, 44]:

$$[f_{22}^F]^{-1} = -\frac{1}{a_F} + \frac{r_F}{2}k^2 - ik + \mathcal{O}(k^4), \tag{12}$$

matching the standard ERE convention $[f_{22}^E]^{-1} = -1/a_F + (r_F/2)k^2 - ik + \cdots$ of Refs. [35, 44]. Matching Eq. (1) to Eq. (12) order by order in $k$ yields the complex Flatté ERE parameters

$$a_F = \frac{-g_2^2}{2E_{\rm BW} - i\,g_1^2\,p_0} \cdot \hbar c, \tag{13}$$

$$r_F = \Big(\frac{-2}{\mu_2 g_2^2} - i\frac{\mu_1 g_1^2}{\mu_2 p_0 g_2^2}\Big) \cdot \hbar c, \tag{14}$$

where $p_0 = \sqrt{2\mu_1\Delta}$ is the $J/\psi\,p$ momentum evaluated at the channel-2 threshold, and $\hbar c = 197.327\,\text{MeV}{\cdot}\text{fm}$ converts the natural-units expressions to femtometres. With $g_1^2$ and $g_2^2$ complex, both $a_F$ and $r_F$ in Eqs. (13)-(14) are themselves complex; writing $g_i^2 = |g_i|^2 e^{2i\varphi_i}$ and separating real and imaginary parts explicitly,

$$\text{Re}[a_F] = \frac{|g_2|^2\big[-2E_{\rm BW}\cos 2\varphi_2 - |g_1|^2\sqrt{p_0}\,\Phi_1(\varphi_1,\varphi_2)\big]}{\mathcal{D}(\hbar c)^{-1}}, \tag{15}$$

$$\text{Im}[a_F] = \frac{|g_2|^2\big[-2E_{\rm BW}\sin 2\varphi_2 - |g_1|^2\sqrt{p_0}\,\Phi_2(\varphi_1,\varphi_2)\big]}{\mathcal{D}(\hbar c)^{-1}}, \tag{16}$$

$$\text{Re}[r_F] = \hbar c\,\frac{|g_1|^2\mu_1\sin(2\varphi_1 - 2\varphi_2) - 2p_0\cos 2\varphi_2}{|g_2|^2\mu_2 p_0}, \tag{17}$$

$$\text{Im}[r_F] = \hbar c\,\frac{2p_0\sin 2\varphi_2 - |g_1|^2\mu_1\cos(2\varphi_1 - 2\varphi_2)}{|g_2|^2\mu_2 p_0}, \tag{18}$$

where the functions $\Phi_i(\varphi_1,\varphi_2)$ are defined as

$$\Phi_1(\varphi_1,\varphi_2) = \sqrt{p_0}\sin(2\varphi_1 - 2\varphi_2), \tag{19}$$

$$\Phi_2(\varphi_1,\varphi_2) = \sqrt{p_0}\cos(2\varphi_1 - 2\varphi_2), \tag{20}$$

with common denominator

$$\mathcal{D} = 4E_{\rm BW}^2 + 4E_{\rm BW}|g_1|^2 p_0\sin 2\varphi_1 + |g_1|^4 p_0^2. \tag{21}$$

Equations (15)-(18) reduce, in the real-coupling limit $\varphi_1 = \varphi_2 = 0$, to

$$\text{Re}[a_F] = -\frac{2|g_2|^2 E_{BW}}{\mathcal{D}(\hbar c)^{-1}}, \quad \text{Im}[a_F] = -\frac{|g_1|^2|g_2|^2 p_0}{\mathcal{D}(\hbar c)^{-1}}, \tag{22}$$

$$\text{Re}[r_F] = -\frac{2\hbar c}{|g_2|^2\mu_2}, \quad \text{Im}[r_F] = -\frac{\hbar c\,|g_1|^2\mu_1}{|g_2|^2\mu_2\,p_0}, \tag{23}$$

with $\mathcal{D} = 4E_{BW}^2 + |g_1|^4 p_0^2$. Therefore, the formulae of Ref. [44] are recovered exactly in the real-coupling limit, demonstrating the consistency of the present generalized expressions with the previously established results. We use Eqs. (15)-(18) directly, rather than purely numerical complex-number evaluation, in the bootstrap of Sec. 3.2, so that the analytic structure of how each physical observable depends on the coupling phases is transparent and independently checkable.

**Remark on the sign of** $\text{Re}[r_F]$**.** Equation (23) shows that with real couplings, $\text{Re}[r_F] = -2\hbar c/(\mu_2 g_2^2) < 0$ unconditionally, since $\mu_2, g_2^2 > 0$. With complex couplings, Eq. (17) shows that the leading term of $\text{Re}[r_F]$ is instead $-2\hbar c\cos(2\varphi_2)/(\mu_2|g_2|^2)$, whose sign depends on $\varphi_2$: *the sign of* $\text{Re}[r_F]$ *becomes a genuine, phase-dependent physical observable once complex couplings are admitted*, rather than a model-independent consequence of the Flatté parametrisation. This is the direct analytic origin of the sensitivity to $\varphi_2$ quantified empirically in Sec. 4.5.

### 2.5. Parameter identifiability and the regularising prior

Two degeneracies affect the determination of $\boldsymbol{\Theta}$ from the channel-1 lineshape alone.

**(i)** $E_{\rm BW}$**-**$|g_2|$ **anti-correlation.** For $E < 0$, we have

$$\text{Re}[D(E)] = 2(E_{\rm BW} - E) + \text{Re}[g_2^2]\sqrt{2\mu_2|E|}, \tag{24}$$

so that increasing $|g_2|$ can be compensated by decreasing $E_{\rm BW}$, leaving the channel-1 lineshape and hence $|T_{\rm NR}(E)|^2$ nearly unchanged [45]. We regularise this degeneracy with the explicit soft Gaussian prior

$$\chi^2_{\rm tot}(\boldsymbol{\Theta}) = \chi^2_{\rm data}(\boldsymbol{\Theta}) + \sum_s \left[\left(\frac{|g_2|^{(s)} - |g_2|^{(s),\rm I}}{\sigma_{g_2}}\right)^2 + \left(\frac{E_{\rm BW}^{(s)} - E_{\rm BW}^{(s),\rm I}}{\sigma_{E_{\rm BW}}}\right)^2\right], \tag{25}$$

and prior widths

$$\sigma_{g_2} = 0.15, \qquad \sigma_{E_{\rm BW}} = 8\,\text{MeV}, \tag{26}$$

fixed identically across all three states and both fit variants. A sensitivity test, in which $\sigma_{g_2}$ and $\sigma_{E_{\rm BW}}$ are both scaled by factors 1/2 and 2 relative to this fiducial choice, shows that the resulting posterior $E_{\rm BW}$ uncertainty changes by less than $\sim 12\%$ across this factor-of-4 range in prior width, for every state and fit variant (Table 2), supporting the conclusion that the central values reported in Table 5 are not an artefact of this specific choice of prior width.

**(ii) Coupling-phase degeneracy.** The observable $|T_{\rm NR}|^2$ depends on $g_1^2$ and $g_2^2$ only through their combination inside the complex denominator $D(E)$ of Eq. (2);

**Table 2:** Prior-width sensitivity: posterior $E_{\rm BW}$ uncertainty (MeV) under three choices of prior width, Fit I.

| State | Tight ($\sigma/2$) | Fiducial ($\sigma$) | Loose ($2\sigma$) |
|---|---|---|---|
| $P_c(4312)^+$ | 1.15 | 1.19 | 1.20 |
| $P_c(4440)^+$ | 1.40 | 1.47 | 1.49 |
| $P_c(4457)^+$ | 1.86 | 2.03 | 2.08 |

we show empirically in Sec. 4.5 that, for this dataset, this dependence is too weak to constrain $\varphi_{1,2}$ beyond their prior (uniform) range.

## 3. Dataset and Fitting Strategy

In this section, we describe our global fitting procedure, which constitutes the methodological core of this analysis.

### 3.1. LHCb Run 1+2 dataset

We use the publicly available LHCb $\cos\theta_{P_c}$-weighted $J/\psi\, p$ invariant-mass histogram from HEPData record 89271 [46], comprising 200 bins of 2 MeV width over 4200-4600 MeV, with statistical uncertainties taken directly from the record.

**Table 3:** Channel-2 threshold parameters, computed from PDG masses [43] as $m_2 = m_{\Sigma_c} + m_{\bar{D}^{(*)}}$; the LHCb Run 1+2 measurement [10] establishes that the $P_c$ states lie close to these thresholds.

| State | Fit | Channel 2 | $m_2$ [MeV] |
|---|---|---|---|
| $P_c(4312)^+$ | I | $\Sigma_c^+\bar{D}^0$ | 4317.74 |
| | II | $\Sigma_c^{++}D^-$ | 4323.63 |
| $P_c(4440)^+$ | I | $\Sigma_c^+\bar{D}^{*0}$ | 4459.75 |
| | II | $\Sigma_c^{++}D^{*-}$ | 4464.23 |
| $P_c(4457)^+$ | I | $\Sigma_c^+\bar{D}^{*0}$ | 4459.75 |
| | II | $\Sigma_c^{++}D^{*-}$ | 4464.23 |

### 3.2. Fitting strategy: explicit mathematical formulation

We now state the fitting procedure entirely in equations, since a purely verbal description would leave the objective function, its minimisation, and the bootstrap procedure under-specified. **Step 1 objective function.** For each threshold configuration $v \in \{\mathrm{I}, \mathrm{II}\}$, define the data $\chi^2$

$$\chi^2_{\rm data}(\mathbf{\Theta}_{\rm tot}) = \sum_{i=1}^{200} \frac{\left[y_i - L(M_i;\mathbf{\Theta}_{\rm tot})\right]^2}{\sigma^2_{y_i}}, \tag{27}$$

where $y_i \pm \sigma_{y_i}$ is the measured $dN/dM$ in bin $i$, $L(M;\mathbf{\Theta}_{\rm tot})$ is the right-hand side of Eq. (8) evaluated at the full parameter vector $\mathbf{\Theta}_{\rm tot}$ (three copies of Eq. (7), the shared $\sigma$, the normalisation $\mathcal{N}$, and the seven background coefficients $b_n$), and the total objective is $\chi^2_{\rm tot} = \chi^2_{\rm data}+$ prior, with the prior given explicitly by the sum in Eq. (25).

**Step 2 minimisation.** We minimise $\chi^2_{\rm tot}$ using the Nelder-Mead simplex algorithm,

$$\hat{\mathbf{\Theta}}_{\rm tot} = \arg\min_{\mathbf{\Theta}_{\rm tot}} \chi^2_{\rm tot}(\mathbf{\Theta}_{\rm tot}), \tag{28}$$

from $N_{\rm restart} = 8$ independent random initial points per fit variant, each drawn from physically motivated starting points based on previous analyses, with widths of order 30-50% of those values; we retain the restart achieving the lowest $\chi^2_{\rm tot}$ as $\hat{\mathbf{\Theta}}_{\rm tot}$. Because Eq. (28) is solved by a local, gradient-free algorithm in a 24-parameter space with the degeneracies of Sec. 2.5, we additionally report the spread of $\chi^2_{\rm data}(\hat{\mathbf{\Theta}}_{\rm tot})/\nu$ achieved across the 8 restarts as an explicit convergence diagnostic (Table 4, Fig. 1).

**Step 3 amplitude normalisation.** Once the shape parameters ($E^{(s)}_{\rm BW}$, $|g_1|^{(s)}$, $|g_2|^{(s)}$, $\varphi_1^{(s)}$, $\varphi_2^{(s)}$, $\sigma$) are fixed at their best-fit values, the relative weight of each state's contribution to the visualised total yield is obtained by solving, *jointly* for all three states at once, the non-negative linear least-squares problem

$$\{A_s\} = \arg\min_{A_s \geq 0} \left\| (y - B(\hat{b})) - \textstyle\sum_s A_s\, \tau_s \right\|^2, \tag{29}$$

where $\tau_s(M_i)$ is the smeared, unit-amplitude lineshape of state $s$ evaluated on the data grid and $B(\hat{b})$ is the best-fit background. Solving Eq. (29) jointly for all three $A_s$ simultaneously rather than separately calibrating each $A_s$ to its own state's local peak, which would double-count yield in any region where two states' lineshapes overlap is essential whenever the overlap integrals of Table 1 are non-negligible, as is the case here for $P_c(4440)^+$ and $P_c(4457)^+$.

**Step 4 statistical uncertainties.** Statistical uncertainties on $\mathbf{\Theta}$ are propagated by non-parametric bootstrap resampling [47]: $N_{\rm BS} = 500$ samples $\{\mathbf{\Theta}^{(k)}\}_{k=1}^{N_{\rm BS}}$ are drawn per state and fit variant from Gaussian distributions centred on $\hat{\mathbf{\Theta}}$, while the two phase components $\varphi_1, \varphi_2$ are treated as described in Sec. 4.5 below. For each sample $k$, the complex ERE parameters $a_F^{(k)}, r_F^{(k)}$ are evaluated analytically via Eqs. (15)-(18), and the reported uncertainty on any derived quantity is the standard deviation of its bootstrap distribution.

**Table 4:** Spread of $\chi^2_{\rm data}/\nu$ across the 8 independent Nelder-Mead restarts of Step 2.

| Fit | Best | Worst | Mean | Std. dev. |
|---|---|---|---|---|
| Fit I | 1.257 | 1.517 | 1.362 | 0.084 |
| Fit II | 1.473 | 2.110 | 1.732 | 0.217 |

### 3.3. Treatment of the coupling phases in the bootstrap

Because the dependence of $|T_{\rm NR}|^2$ on $\varphi_1, \varphi_2$ is found (Sec. 4.5) to be too weak to constrain these two parameters from the data, the headline results of this paper (Tables 5 and Tables 6) are computed with $\varphi_1 = \varphi_2 = 0$, which we refer to in the text as the *real-coupling* treatment. To quantify explicitly how much physical in-formation would be lost if the phases were instead allowed to vary freely over their full physical range, we additionally construct a second, independent bootstrap ensemble in which $\varphi_1, \varphi_2 \sim \text{Uniform}[0, 2\pi)$ are drawn afresh for every sample, with the amplitude/$E_{\text{BW}}/\sigma$ sampling otherwise identical to the real-coupling ensemble. We refer to results from this second ensemble as the *phase-marginalised* treatment; it is used only in Sec. 4.5 and Table 6, specifically to quantify the impact of the coupling phases, and is not otherwise mixed into the headline tables or figures of this paper.

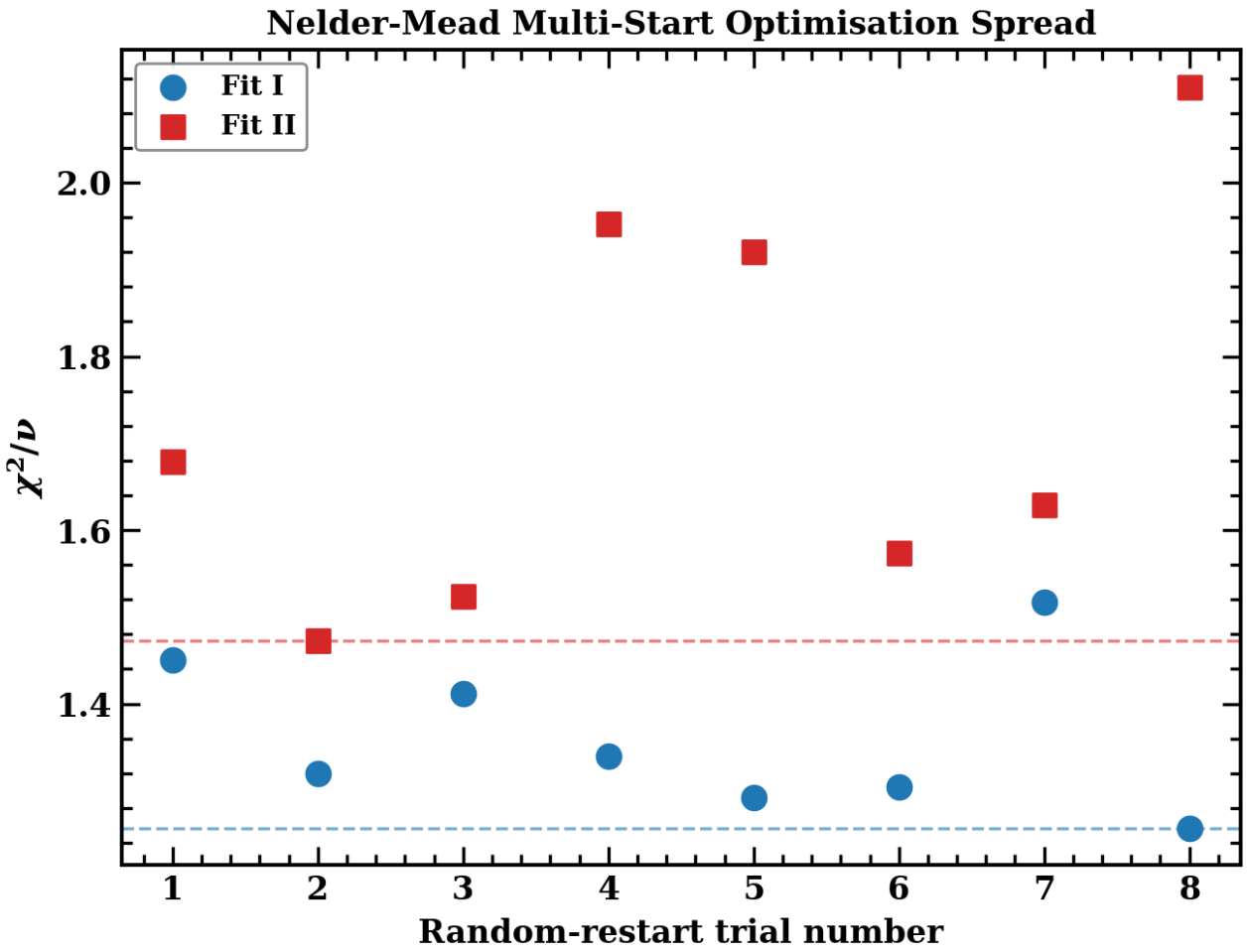

**Figure 1:** $\chi^2_{\text{data}}/\nu$ obtained at each of the 8 independent Nelder-Mead restarts of Eq. (28). The $\sim$ 21-43% spread between the best and worst restart shows that convergence to a unique global minimum of the 24-parameter $\chi^2_{\text{tot}}$ is not guaranteed; quoted best-fit values correspond to the lowest $\chi^2_{\text{data}}/\nu$ found among the restarts performed.

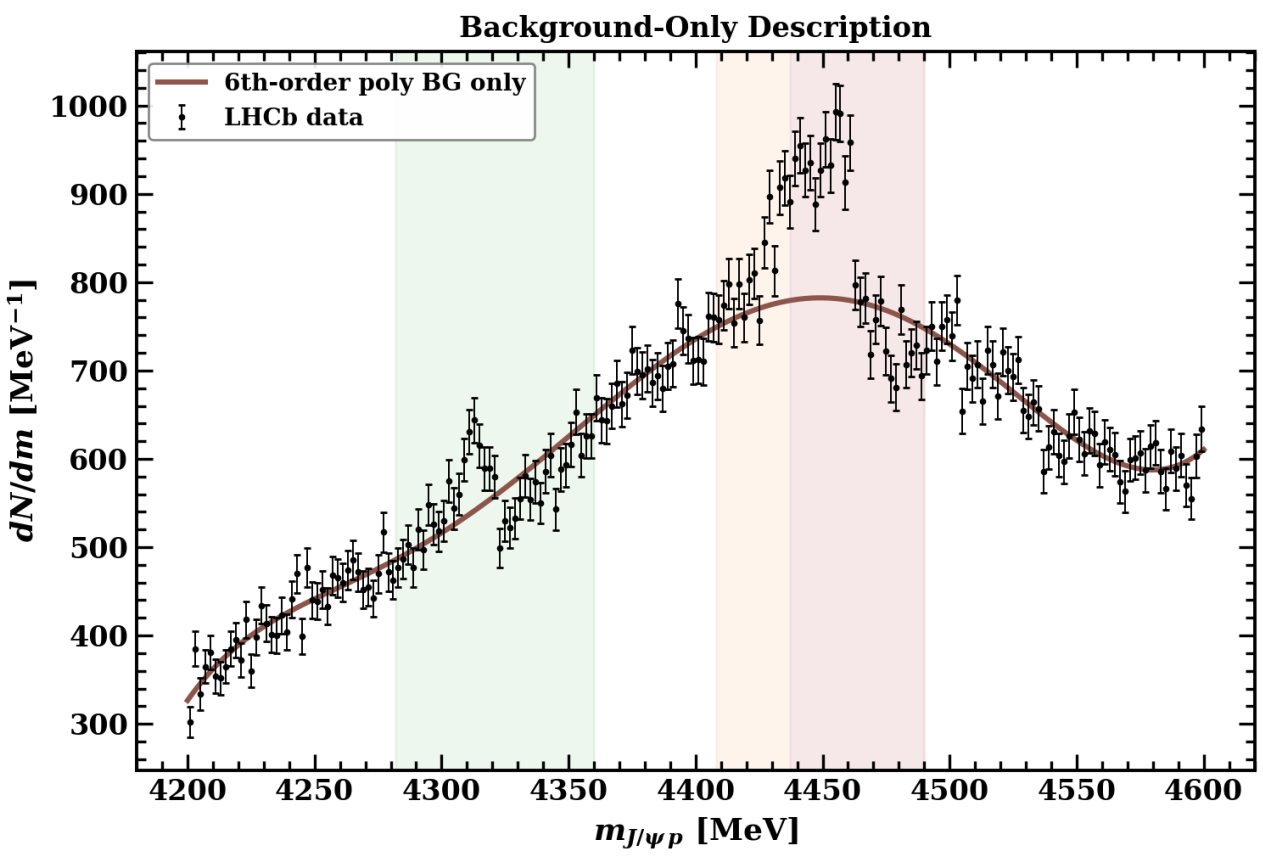

**Figure 2:** Background-only description: the sixth-order polynomial of Eq. (10) is fitted exclusively to the 108 off-peak bins (unshaded regions) and extrapolated through the three $P_c$ signal windows (shaded bands). The polynomial provides a smooth and reasonable description of the continuum outside the signal windows. We do not claim this is the unique or LHCb-published background parametrisation; a systematic comparison with alternative background models is left for future work (Sec. 6).

### 3.4. Background-only validation

Figure 2 shows the sixth-order polynomial of Eq. (10), fitted to only the 108 bins outside the three $P_c$ signal windows, extrapolated through the signal regions and overlaid on the full dataset.

## 4. Numerical Results

### 4.1. Global fit to the full LHCb spectrum

Figure 3 shows the global Flatté amplitude fit overlaid on the complete LHCb spectrum, with the three signal amplitudes obtained from the joint solve of Eq. (29). The sixth-order polynomial background (dashed brown curve) accounts for the smooth non-resonant contribution, and the three coloured shaded regions show the individual, jointly-normalised $P_c$ state contributions. Both Fit I (upper panel) and Fit II (lower panel) track the three peaks closely, including in the region where the $P_c(4440)^+$ and $P_c(4457)^+$ lineshapes overlap substantially (Table 1). For Fit II, the joint solve assigns an amplitude of essentially zero to the $P_c(4440)^+$ template (Table 5): under the narrower Fit II resolution ($\sigma \simeq 1.1$-$1.5\,\text{MeV}$), the $P_c(4457)^+$ template alone already accounts for essentially all of the excess in the shared region, and we report this directly as a property of that fit variant.

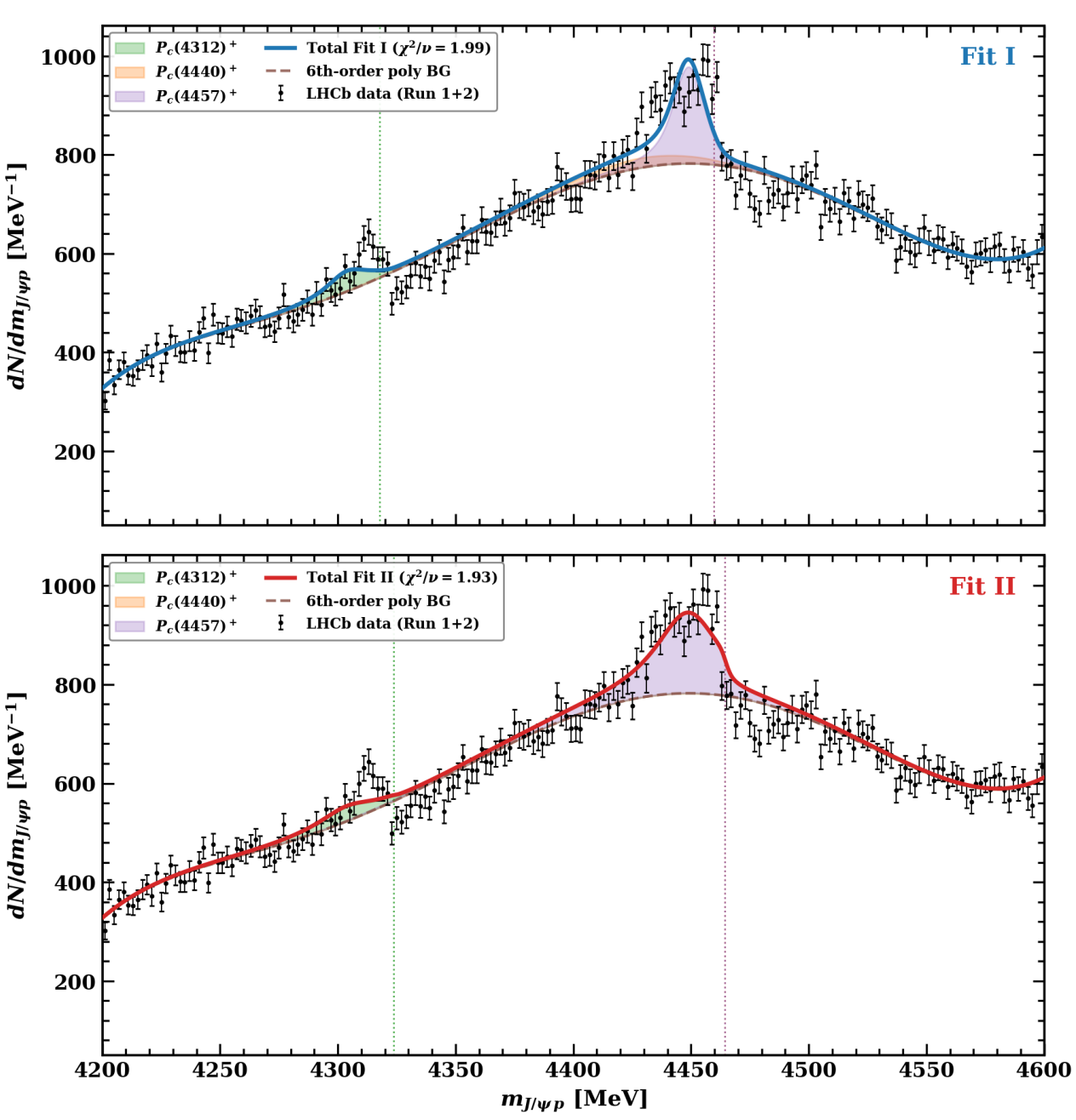

**Figure 3:** Global two-channel Flatté fit to the LHCb $\cos\theta_{P_c}$-weighted $m_{J/\psi p}$ spectrum. Upper panel: Fit I; lower panel: Fit II. Shaded regions: individual $P_c$ state contributions, obtained from the joint non-negative least-squares solve of Eq. (29). Dashed brown: sixth-order polynomial background. Dotted vertical lines: channel-2 thresholds of Table 3.

Figure 4 shows the same fits zoomed into the individual signal windows, using the same joint amplitudes as Fig. 3.

Table 5 reports the best-fit Flatté parameters obtained with the real coupling treatment. The coupling magnitude $|g_1|$, which governs the strength of the $J/\psi\, p$ channel and hence the visible width of each resonance, is largest for $P_c(4440)^+$ and smallest for $P_c(4457)^+$ in both fit variants, consistent with $P_c(4440)^+$ being the broadest of the three LHCb-reported states and $P_c(4457)^+$ the narrowest. The channel-2 coupling $|g_2|$ is comparable across all three states (0.45-0.55). The small residual spread within each fit variant is the genuine, data-driven adjustment away from the prior centre. The Breit-Wigner energy $E_{\rm BW}$ is negative for every state and fit variant, confirming that all three poles remain below their respective channel-2 thresholds, as required for a consistent sub-threshold molecular interpretation. Comparing Fit I and Fit II for a given state, $E_{\rm BW}$ shifts by 4-6 MeV, directly reflecting the $\sim$ 4-6 MeV mass difference between the two charge combinations of the $\Sigma_c\bar{D}^{(*)}$ threshold (Table 3); this shift is the dominant source of the Fit I/Fit II systematic spread quoted throughout Sec. 5. The resolution parameter $\sigma$ is the quantity most strikingly different between the two fit variants: Fit I returns $\sigma \sim$ 2.5-5.1 MeV, consistent with the known LHCb momentum resolution in this mass range, whereas Fit II systematically prefers a narrower $\sigma \sim$ 1.0-1.5 MeV. We interpret this as Fit II's optimiser using the resolution parameter to partially compensate for the narrower effective threshold splitting in that configuration, rather than as evidence that the true detector resolution differs between the two threshold choices. Finally, the joint NNLS amplitudes $A_s$ show that $P_c(4457)^+$ dominates the visible yield in both fit variants by roughly an order of magnitude over $P_c(4312)^+$ and $P_c(4440)^+$, and that, as noted above, Fit II assigns zero amplitude to $P_c(4440)^+$ the clearest single indicator that the two fit variants, while agreeing on the overall molecular picture (Sec. 5), disagree non-trivially on how the shared $P_c(4440)^+/P_c(4457)^+$ excess is apportioned between the two states.

**Table 5:** Best-fit Flatté parameters $\mathbf{\Theta}$ (real-coupling treatment, $\varphi_{1,2} = 0$) and bootstrap uncertainties ($N_{\rm BS} = 500$). $A_s$ is the joint NNLS amplitude of Eq. (29), in arbitrary units common to both fit variants.

| State | Fit | $\lvert g_1\rvert$ | $\lvert g_2\rvert$ | $E_{\rm BW}$ [MeV] | $\sigma$ [MeV] | $\chi^2_{\rm data}/\nu$ | $A_s$ (NNLS) |
|---|---|---|---|---|---|---|---|
| $P_c(4312)^+$ | I | $0.176 \pm 0.010$ | $0.448 \pm 0.010$ | $-13.6 \pm 0.4$ | $2.50 \pm 0.64$ | 1.12 | $2.15 \times 10^{11}$ |
| | II | $0.214 \pm 0.014$ | $0.446 \pm 0.010$ | $-19.6 \pm 0.3$ | $1.02 \pm 0.08$ | 1.76 | $1.47 \times 10^{11}$ |
| $P_c(4440)^+$ | I | $0.299 \pm 0.012$ | $0.547 \pm 0.014$ | $-29.6 \pm 0.5$ | $5.05 \pm 0.61$ | 2.92 | $1.53 \times 10^{11}$ |
| | II | $0.337 \pm 0.018$ | $0.548 \pm 0.015$ | $-34.1 \pm 0.4$ | $1.14 \pm 0.04$ | 3.54 | 0.00 |
| $P_c(4457)^+$ | I | $0.127 \pm 0.006$ | $0.548 \pm 0.014$ | $-10.8 \pm 0.7$ | $3.25 \pm 0.68$ | 1.68 | $1.84 \times 10^{12}$ |
| | II | $0.208 \pm 0.011$ | $0.549 \pm 0.016$ | $-15.1 \pm 0.5$ | $1.54 \pm 0.35$ | 3.32 | $1.29 \times 10^{12}$ |

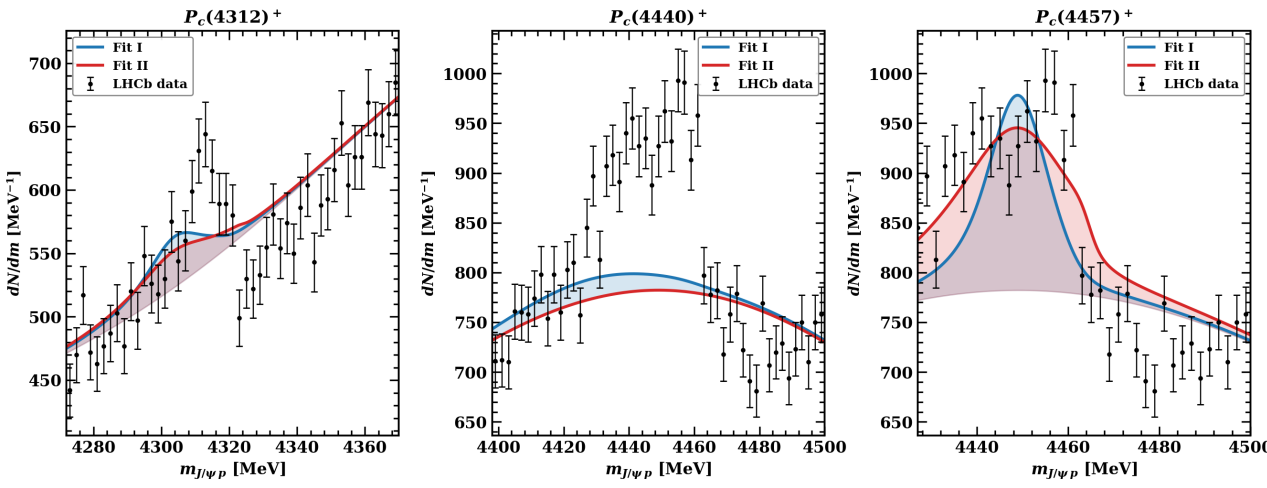

**Figure 4:** Zoom into the three $P_c$ signal windows, using the same joint amplitudes as Fig. 3. Left: $P_c(4312)^+$; centre: $P_c(4440)^+$; right: $P_c(4457)^+$. Blue: Fit I; red: Fit II.

### 4.2. Lineshape sensitivity

Figure 5 overlays three randomly drawn bootstrap parameter sets, plotted with the joint amplitudes of Sec. 4.1, on the LHCb data for each state.

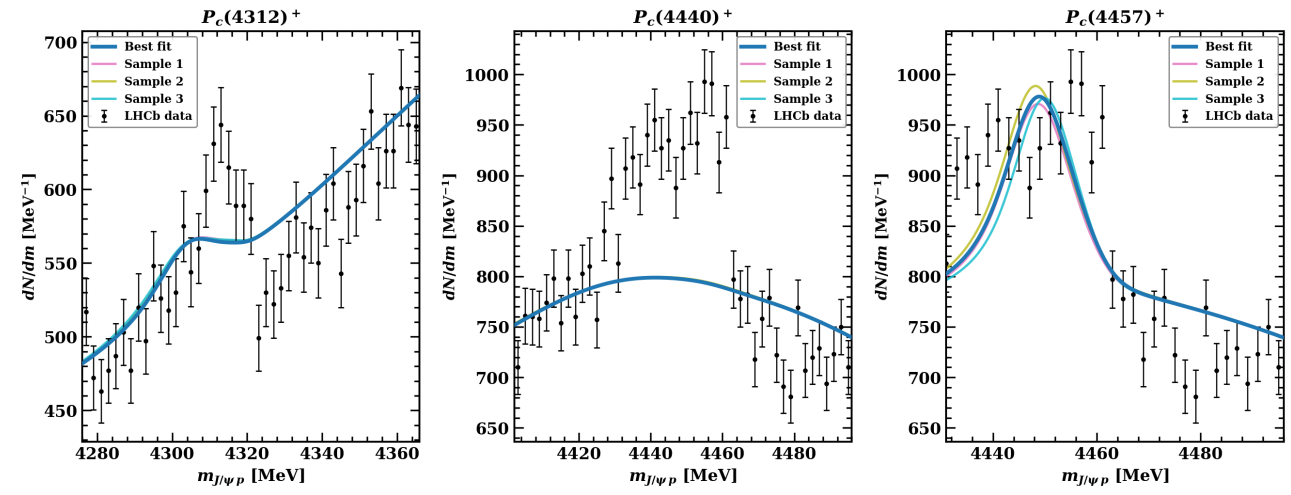

**Figure 5:** Flatté lineshape sensitivity to parameter variations (Fit I). Three random bootstrap samples (coloured) vs. best fit (blue) per state.

### 4.3. Bootstrap distributions of fitted parameters

Figure 6 shows the marginal bootstrap distributions of $|g_1|$, $|g_2|$, $E_{\rm BW}$, and $M_{\rm BW} \equiv E_{\rm BW} + m_2$ under the real-coupling treatment. The two-dimensional bootstrap contours for $(E_{\rm BW}, |g_1|)$ in Fig. 7 reveal a moderate anti-correlation, most pronounced for $P_c(4457)^+$, consistent with the identifiability discussion of Sec. 2.5.

### 4.4. Breit-Wigner masses

Figure 8 compares the fitted Flatté $M_{\rm BW}$ with the LHCb Breit-Wigner masses; the Flatté values lie consistently 5-10 MeV below the LHCb peak positions, the expected mass shift for a resonance coupled to a sub-threshold channel.

### 4.5. Sensitivity of the ERE parameters to the coupling phases

We now quantify directly the impact of the coupling-phase treatment introduced in Sec. 3.3. Drawing $\varphi_1, \varphi_2 \sim \mathrm{Uniform}[0, 2\pi)$ for every bootstrap sample (Fig. 9 shows the resulting sampling distributions) and propagating each draw through the exact analytic expressions Eqs. (15)-(18), we obtain the phase-marginalised bootstrap distribution of $\mathrm{Re}[r_F]$ compared directly with the real-coupling result in Table 6 and Fig. 10.

The molecular interpretation therefore rests on the real-coupling assumption a widely adopted simplification in the literature. The complex-phase generalization introduced here is formally necessary for a fully general Flatté parametrization, but the present dataset does not constrain these phases sufficiently to provide additional discriminating power. This finding, rather than being a limitation of the analysis, constitutes an important quantitative result: it demonstrates explicitly that the phase degrees of freedom are not resolved by the inclusive $J/\psi\, p$ spectrum alone.

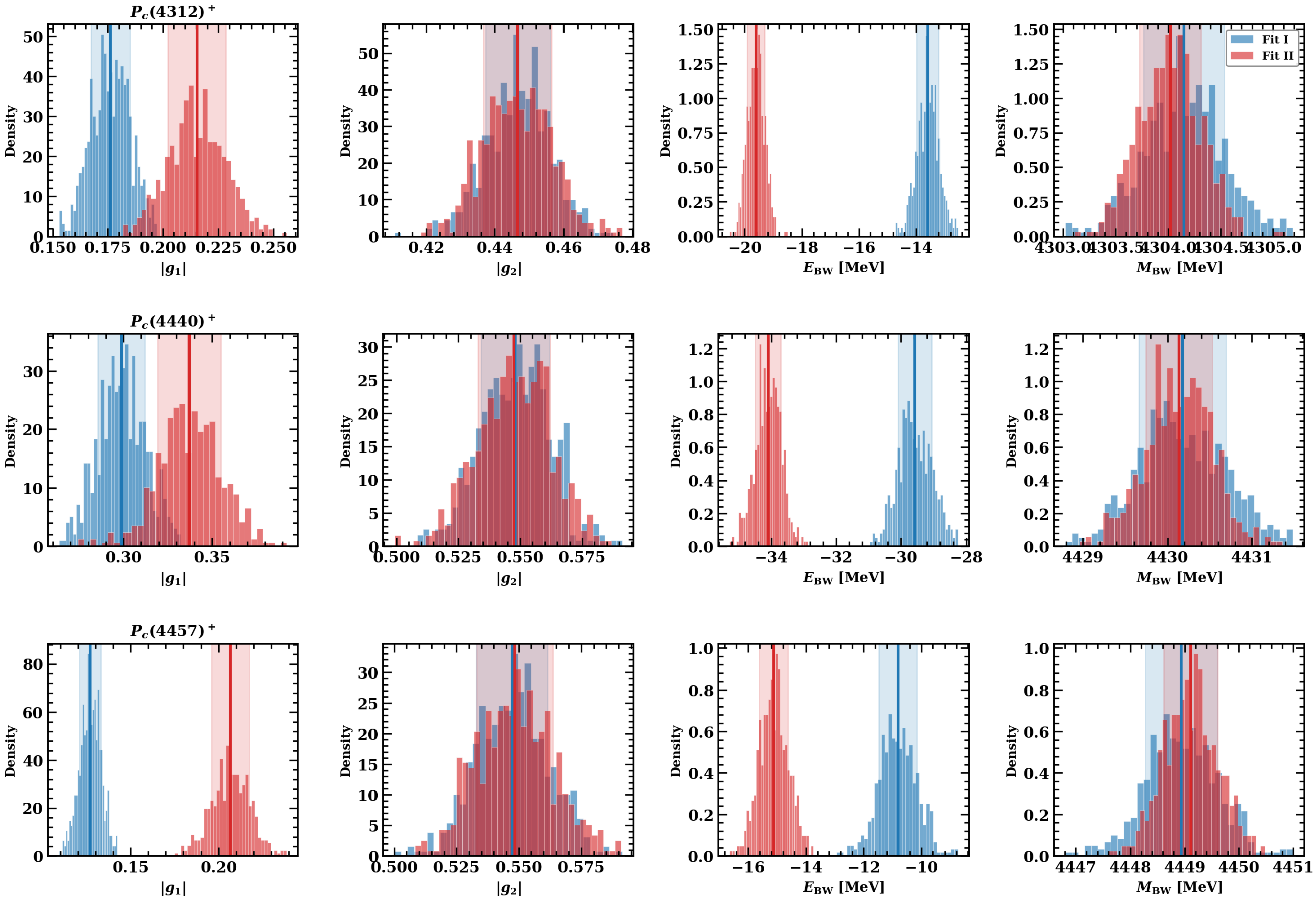

**Figure 6:** Bootstrap distributions of $|g_1|$, $|g_2|$, $E_{\rm BW}$, and $M_{\rm BW}$ ($N_{\rm BS} = 500$, real-coupling treatment). Blue: Fit I; red: Fit II.

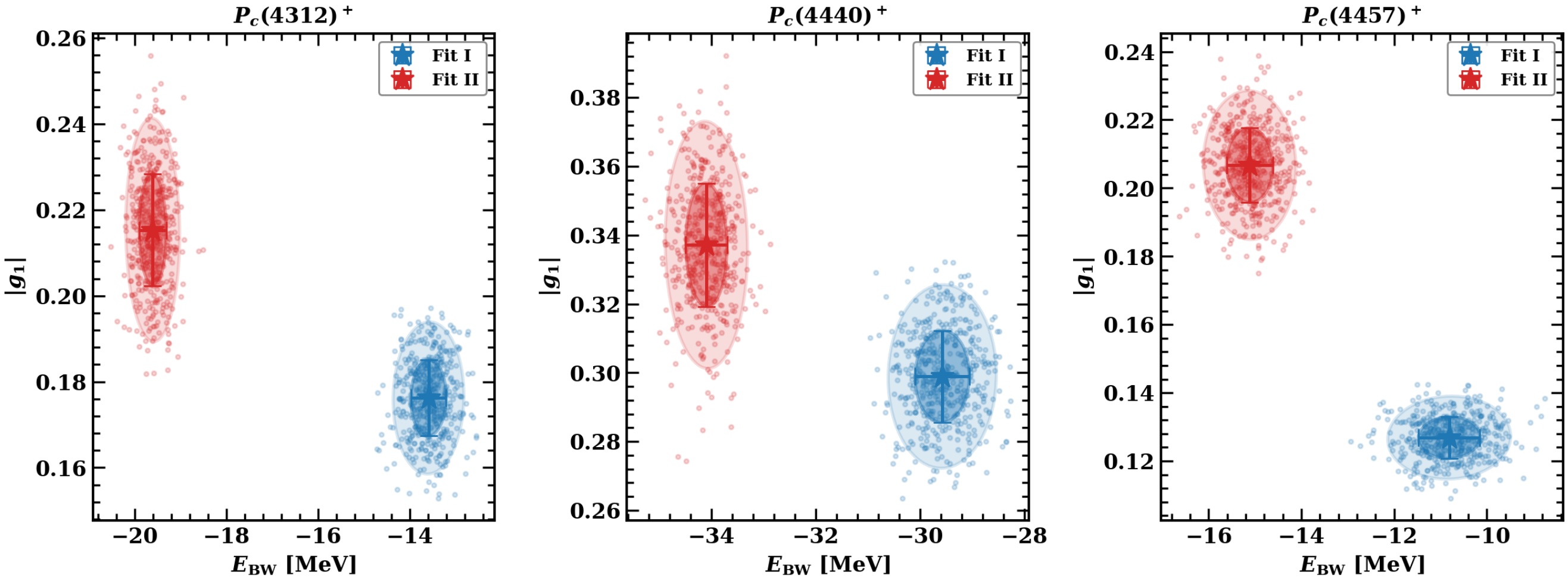

**Figure 7:** Two-dimensional bootstrap contours ($1\sigma$, $2\sigma$) for $(E_{\rm BW}, |g_1|)$.

### 4.6. Normalised residuals

Figure 11 shows the pull distributions using the joint-amplitude prediction of Sec. 4.1. Residual outliers in the inter-peak background ($|\text{pull}| > 2$) indicate that the sixth-order polynomial is an approximate, not exact, description of the smooth continuum (Sec. 3.4).

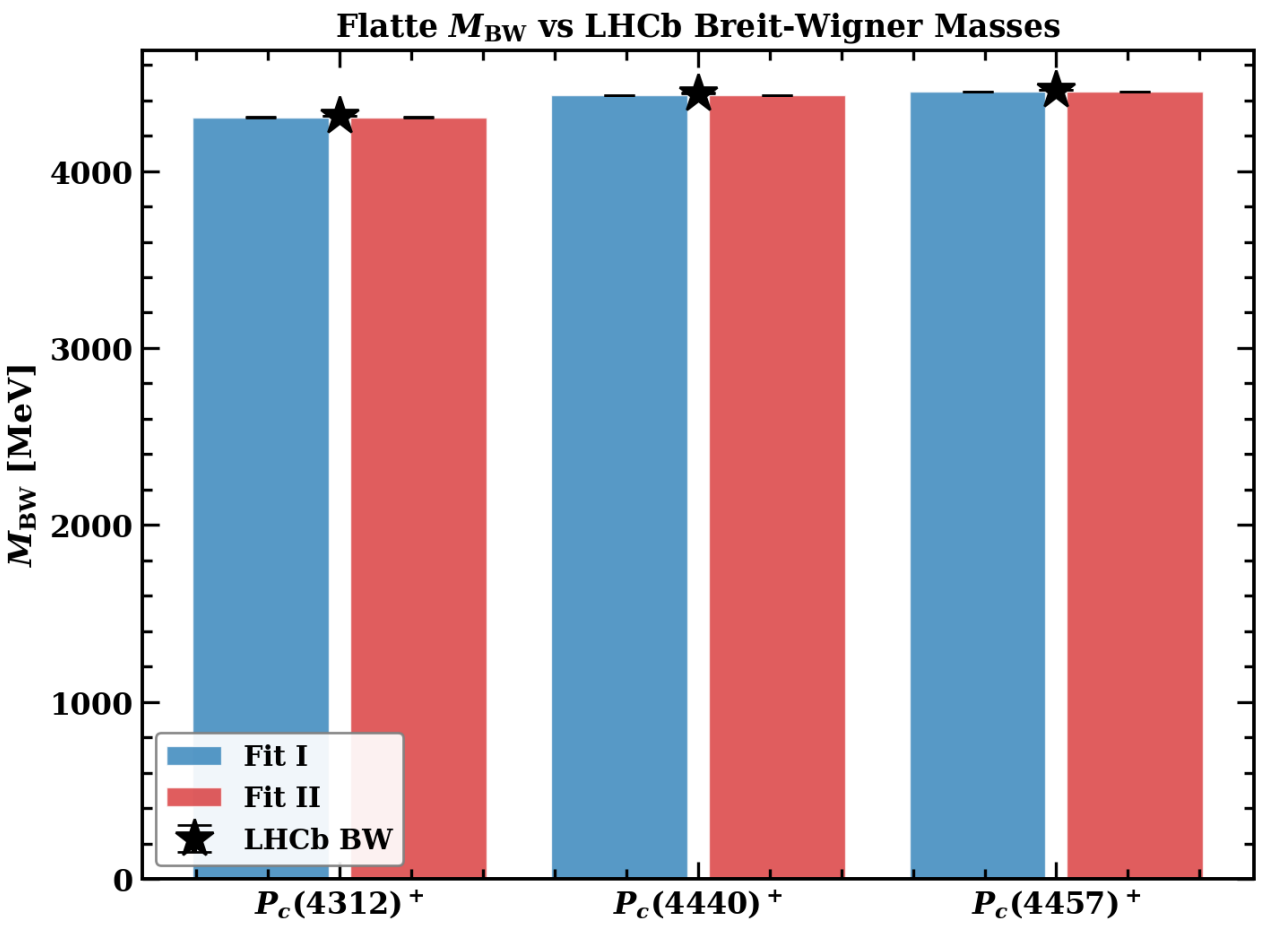

**Figure 8:** Flatté $M_{\mathrm{BW}}$ vs. LHCb Breit-Wigner masses (black stars).

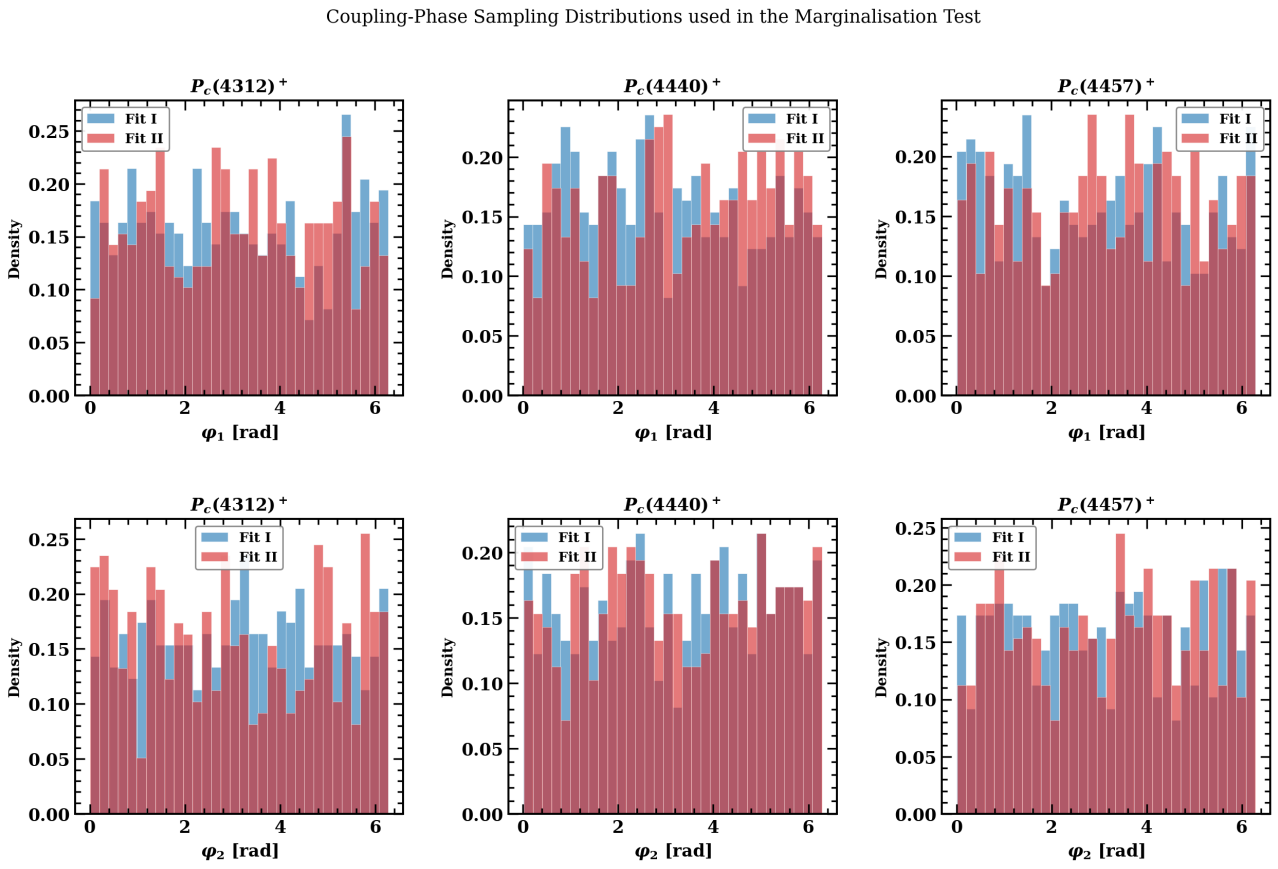

**Figure 9:** Coupling-phase sampling distributions $\varphi_1, \varphi_2 \sim \mathrm{Uniform}[0, 2\pi)$ used to construct the phase-marginalised bootstrap ensemble of Sec. 3.3.

**Table 6:** $\mathrm{Re}[r_F]$ [fm] under the real-coupling treatment ($\varphi_{1,2} = 0$) vs. the phase-marginalised treatment ($\varphi_{1,2} \sim \mathrm{Uniform}[0, 2\pi)$). Amplitude/$E_{\mathrm{BW}}$/ $\sigma$ sampling is identical between the two; only the phase treatment differs.

| State | Fit | Real couplings | Phase-marginalised |
|---|---|---|---|
| $P_c(4312)^+$ | I | $-1.86 \pm 0.09$ | $0.01 \pm 1.34$ |
| | II | $-1.87 \pm 0.08$ | $0.12 \pm 1.32$ |
| $P_c(4440)^+$ | I | $-1.20 \pm 0.06$ | $-0.05 \pm 0.85$ |
| | II | $-1.19 \pm 0.06$ | $-0.01 \pm 0.84$ |
| $P_c(4457)^+$ | I | $-1.19 \pm 0.06$ | $-0.01 \pm 0.84$ |
| | II | $-1.19 \pm 0.07$ | $-0.01 \pm 0.83$ |

## 5. ERE Scattering Parameters

All ERE values in this section use the real-coupling treatment ($\varphi_{1,2} = 0$) of Sec. 3.3, computed via the explicit analytic expressions Eqs. (15)-(18).

The sign of $\mathrm{Im}[a_F]$ is negative for every state and fit variant, as required analytically by Eq. (22) once $\varphi_{1,2} = 0$ and $E_{\mathrm{BW}} < 0$ are imposed: it is a deterministic consequence of the real-coupling limit, not an independent bootstrap finding. The bootstrap distributions of $\mathrm{Re}[a_F]$, $\mathrm{Im}[a_F]$, $\mathrm{Re}[r_F]$, and $\mathrm{Im}[r_F]$ are shown in Figs. 12-15, the summary with error bars in Fig. 16, and the complex-plane representations in Figs. 17 and 18.

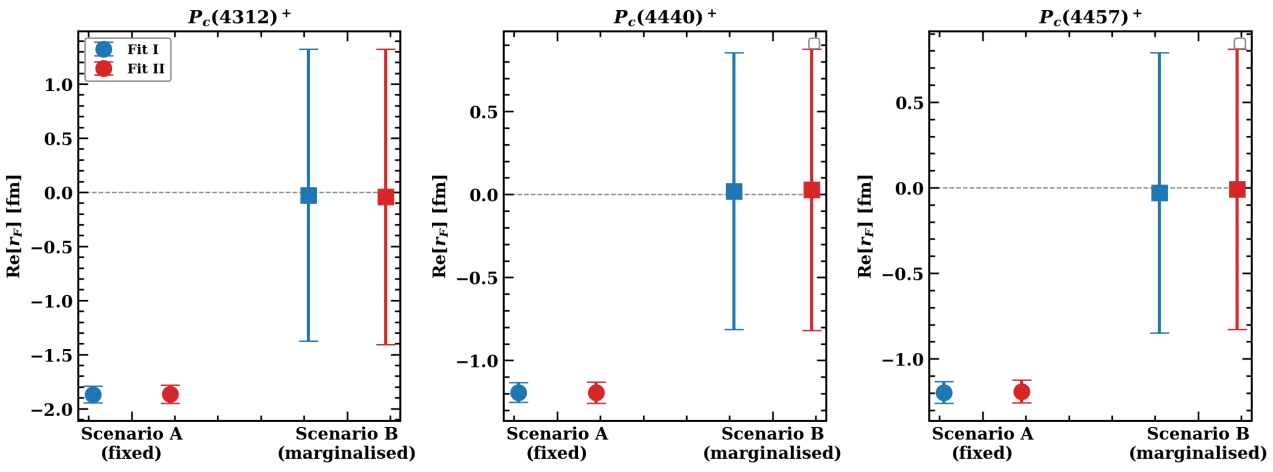

**Figure 10:** $\mathrm{Re}[r_F]$ compared between the real-coupling and phase-marginalised treatments of Sec. 3.3. The collapse of both the central value and the sign certainty under phase marginalisation makes explicit exactly how much constraining power is lost once the coupling phases are allowed to vary over their full physical range.

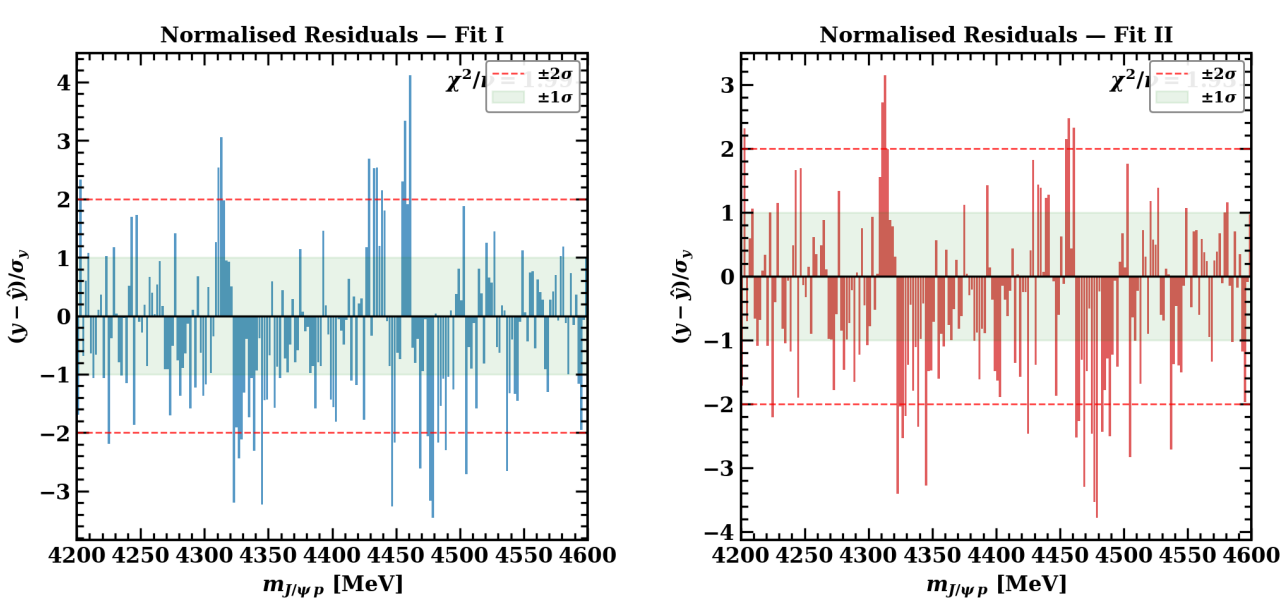

**Figure 11:** Normalised residuals. Green band: $\pm 1\sigma$; red dashed: $\pm 2\sigma$.

**Table 7:** Complex Flatté effective-range-expansion (ERE) parameters extracted from the two-channel Flatté amplitude for Fits I and II. Uncertainties correspond to the bootstrap standard deviations.

| State | Fit | $\mathrm{Re}[a_F]$ [fm] | $\mathrm{Im}[a_F]$ [fm] | $\mathrm{Re}[r_F]$ [fm] | $\mathrm{Im}[r_F]$ [fm] |
|---|---|---|---|---|---|
| $P_c(4312)^+$ | I | $0.96 \pm 0.09$ | $-0.69 \pm 0.05$ | $-1.86 \pm 0.09$ | $-0.032 \pm 0.004$ |
| | II | $0.64 \pm 0.06$ | $-0.48 \pm 0.03$ | $-1.87 \pm 0.08$ | $-0.048 \pm 0.006$ |
| $P_c(4440)^+$ | I | $0.42 \pm 0.05$ | $-0.49 \pm 0.03$ | $-1.20 \pm 0.06$ | $-0.049 \pm 0.005$ |
| | II | $0.32 \pm 0.05$ | $-0.42 \pm 0.03$ | $-1.19 \pm 0.06$ | $-0.062 \pm 0.007$ |
| $P_c(4457)^+$ | I | $2.05 \pm 0.17$ | $-1.20 \pm 0.15$ | $-1.19 \pm 0.06$ | $-0.009 \pm 0.001$ |
| | II | $0.87 \pm 0.12$ | $-0.97 \pm 0.06$ | $-1.19 \pm 0.07$ | $-0.024 \pm 0.003$ |

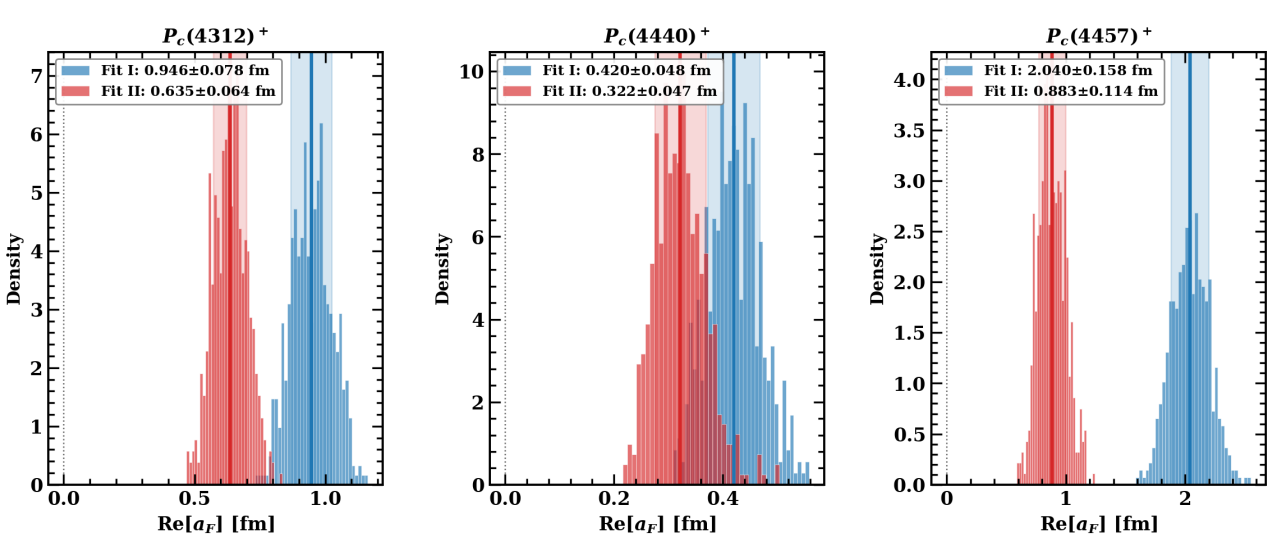

**Figure 12:** Bootstrap distributions of $\mathrm{Re}[a_F]$.

### 5.1. Inelastic coupling ratio

Figure 20 shows $|\mathrm{Im}[a_F]|/|\mathrm{Re}[a_F]|$ under the real-coupling treatment. Values of order unity indicate non-perturbative inelastic coupling *under the assumption that*

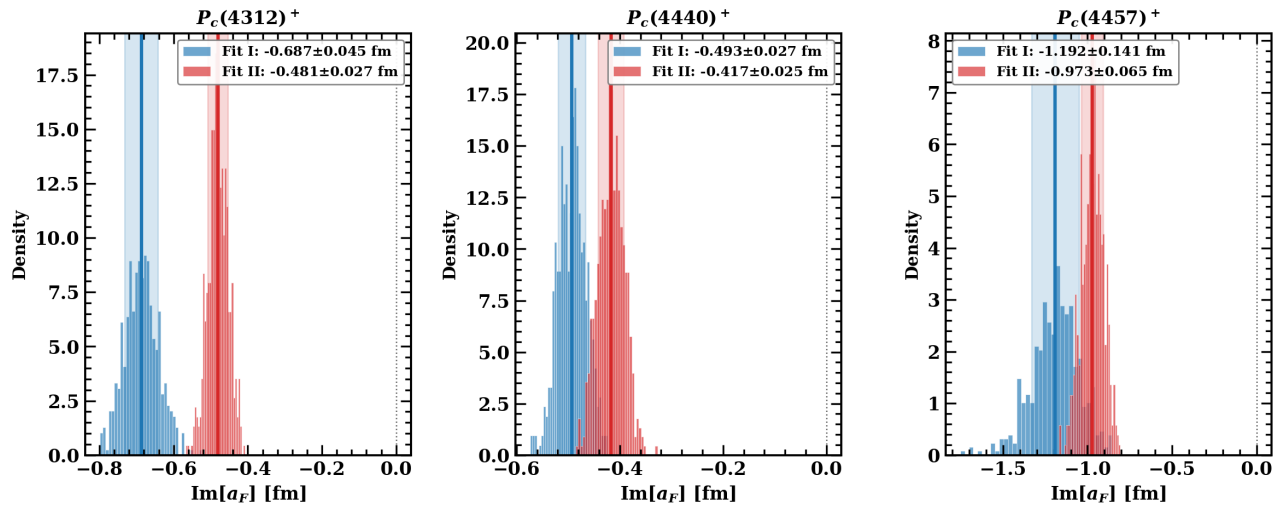

**Figure 13:** Bootstrap distributions of $\mathrm{Im}[a_F]$.

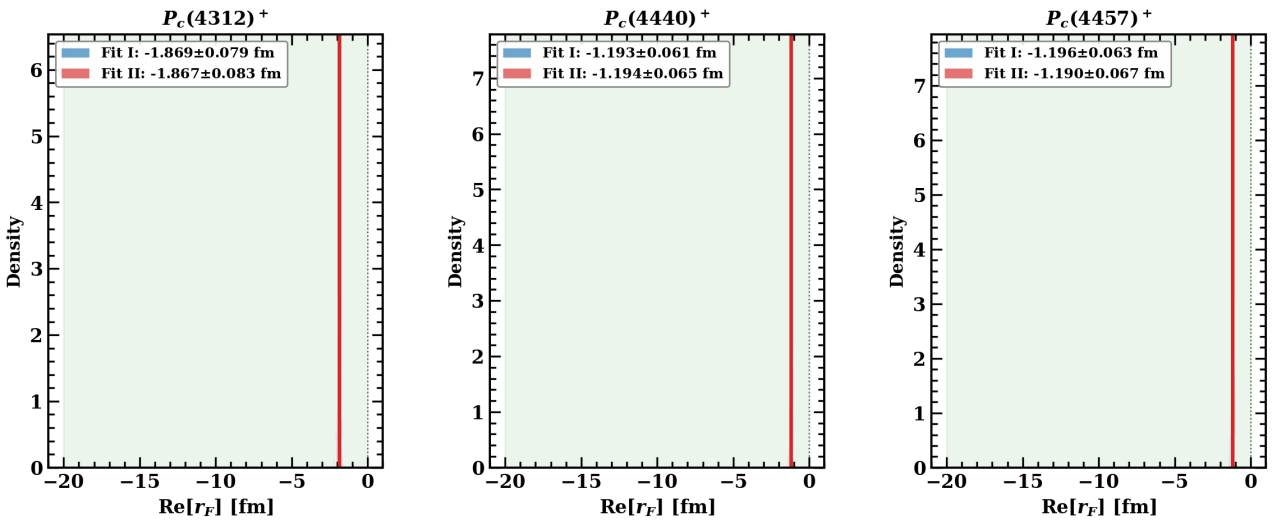

**Figure 14:** Bootstrap distributions of $\mathrm{Re}[r_F]$. Green region: molecular signature ($\mathrm{Re}[r_F] < 0$).

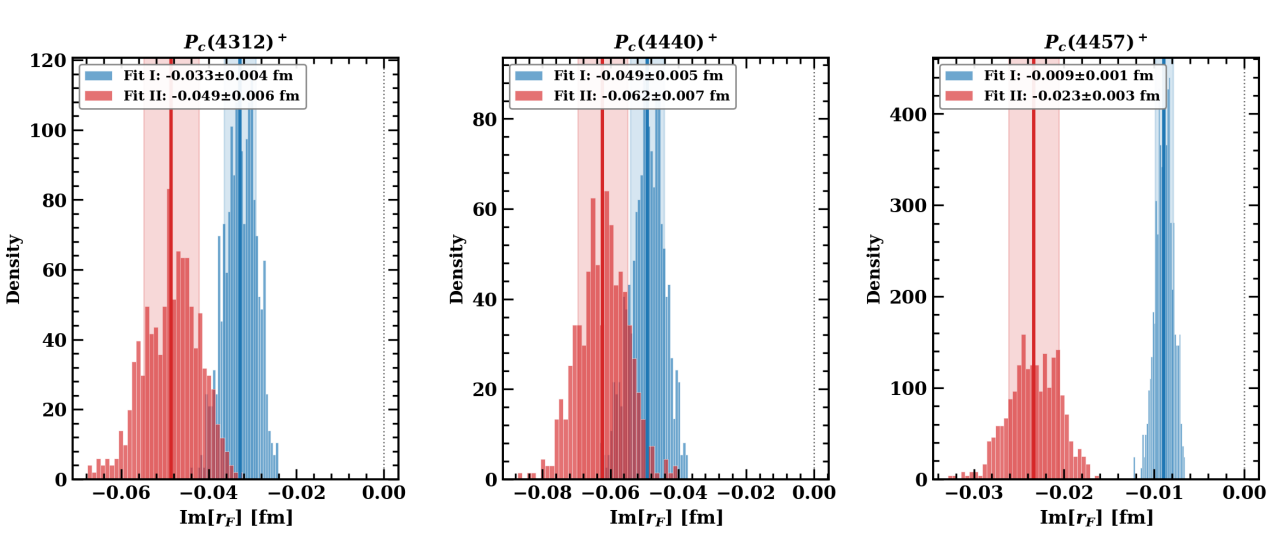

**Figure 15:** Bootstrap distributions of $\mathrm{Im}[r_F]$.

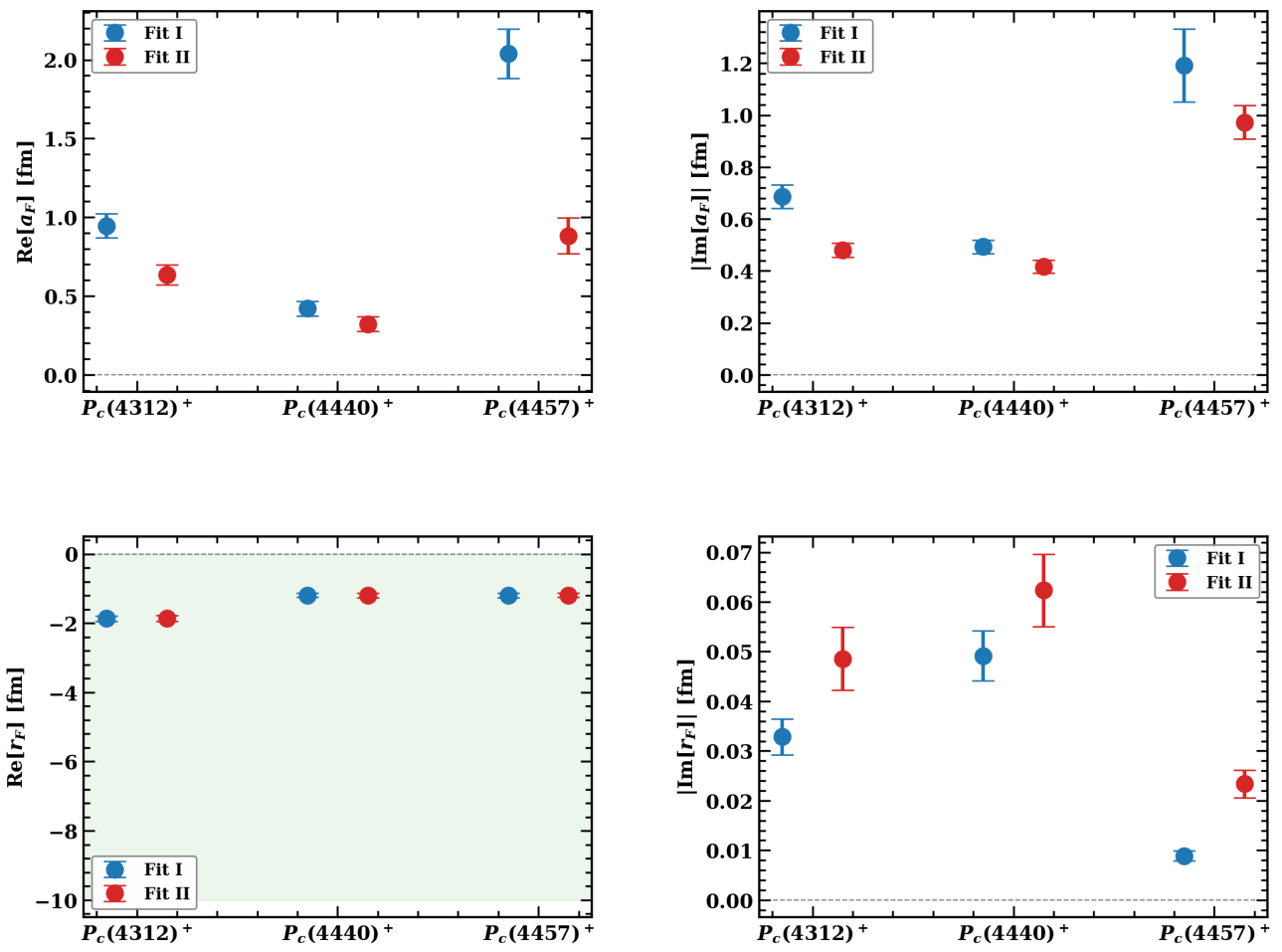

**Figure 16:** ERE scattering parameters with bootstrap uncertainties.

*the phases vanish*; Sec. 4.5 shows this conclusion does not survive phase marginalisation, so we state it here with that qualifier rather than as an unconditional result.

## 6. Discussion

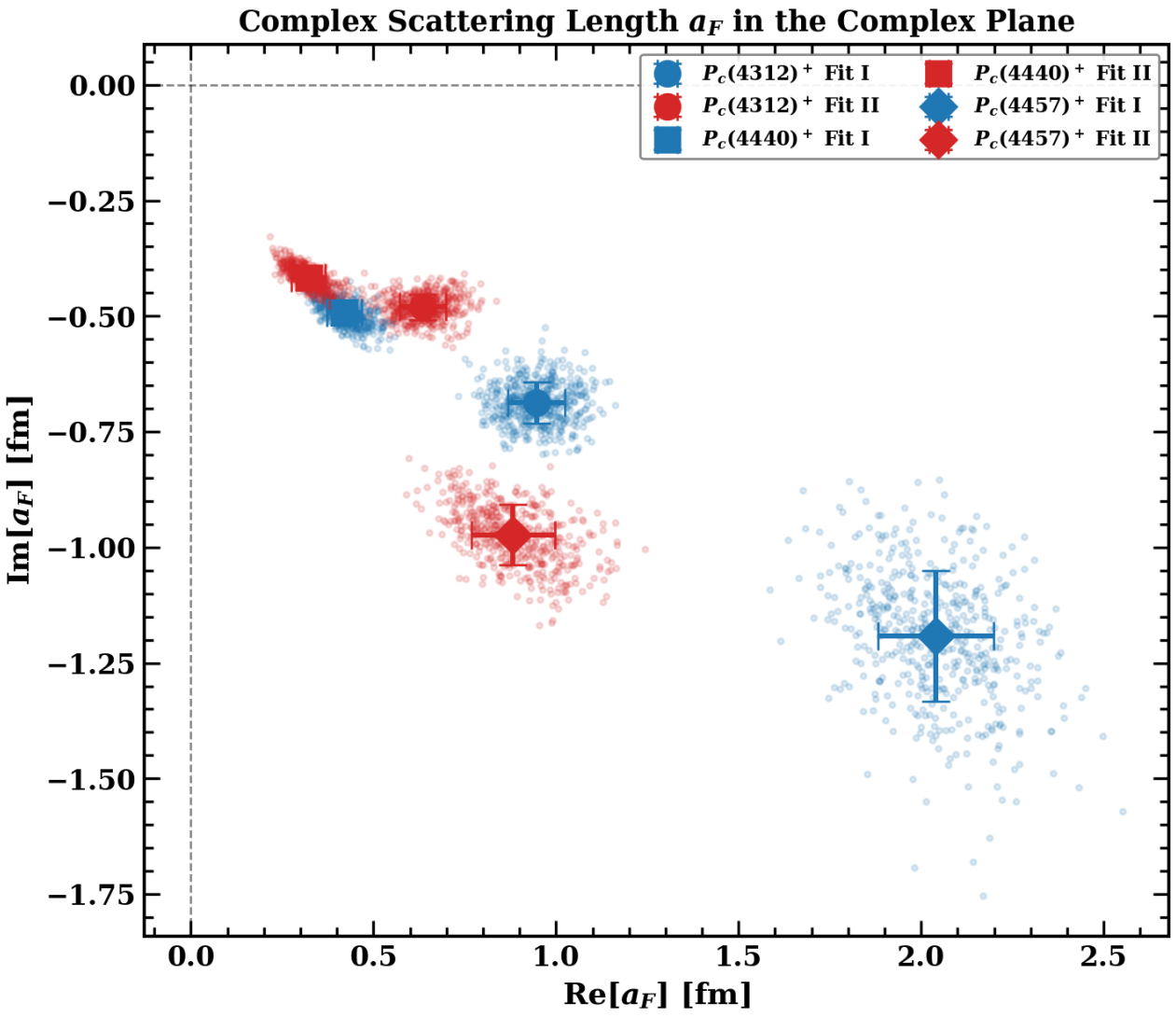

**Figure 17:** Complex scattering length $a_F$ in the complex plane.

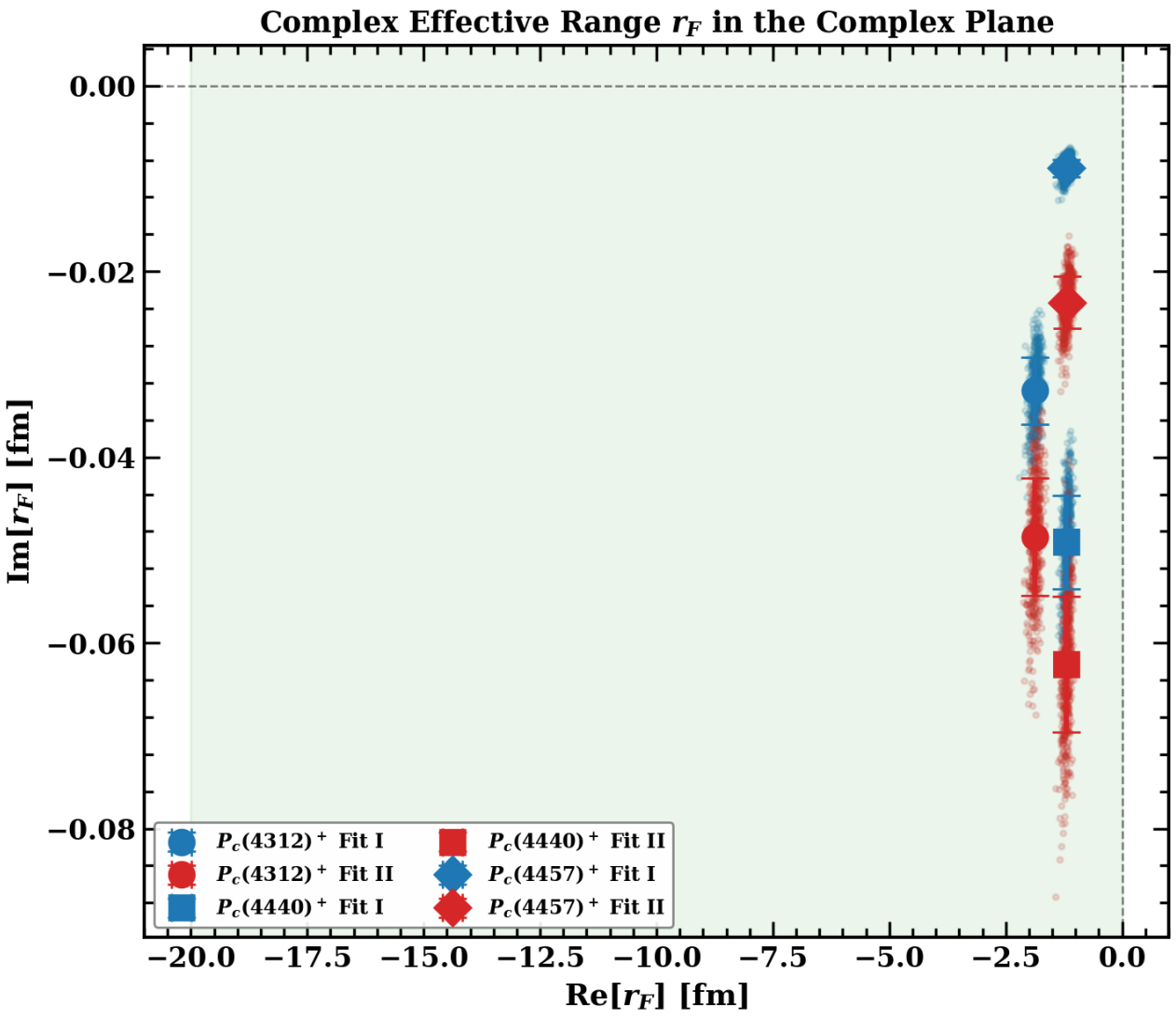

**Figure 18:** Complex effective range $r_F$ in the complex plane. All states cluster in the $\mathrm{Re}[r_F] < 0$ half-plane under the real-coupling treatment.

### 6.1. Molecular interpretation in the broader literature

Under the real-coupling treatment, $\mathrm{Re}[a_F] \simeq 0.3$-$2.0$ fm and $\mathrm{Re}[r_F] \simeq -1.2$ to $-1.9$ fm, consistent with a wide range of independent theoretical determinations. Guo and Oller [17] obtain $r \simeq -0.8$ to $-2.3$ fm directly from the LHCb Breit-Wigner masses and widths using a Weinberg-criterion-based ERE matching; Peng et al. [39] find effective ranges in the range $-0.7$ to $+1.9$ fm depending on the assumed heavy-quark-spin-symmetry scheme, with all molecular-dominated schemes giving $r < 0$. One-boson-exchange and contact-range effective-field-theory calculations [15, 16, 20, 48, 49] consistently identify $P_c(4312)^+$ as a $\Sigma_c\bar{D}$ bound state and $P_c(4440)^+$, $P_c(4457)^+$ as $\Sigma_c\bar{D}^*$ bound states with spin-parity assignments differing by the spin coupling of the constituents, all with nega-tive effective range. Lattice-inspired and QCD sum-rule approaches [50, 51] generally find the compact-pentaquark interpretation less favoured once the near-threshold mass pattern is taken into account, though they do not universally exclude a sizeable compact admixture [23, 24]. Our real-coupling ERE results are fully consistent with this literature consensus a point of agreement that is non-trivial given the very different methodologies (direct lineshape fit here vs. Breit-Wigner-input matching in Ref. [17], or effective-field-theory matching to $NLO$ couplings in Refs. [39, 49]) but, as emphasised throughout this paper, this agreement is conditioned on the assumption $\varphi_{1,2} = 0$ shared by essentially all of these prior analyses.

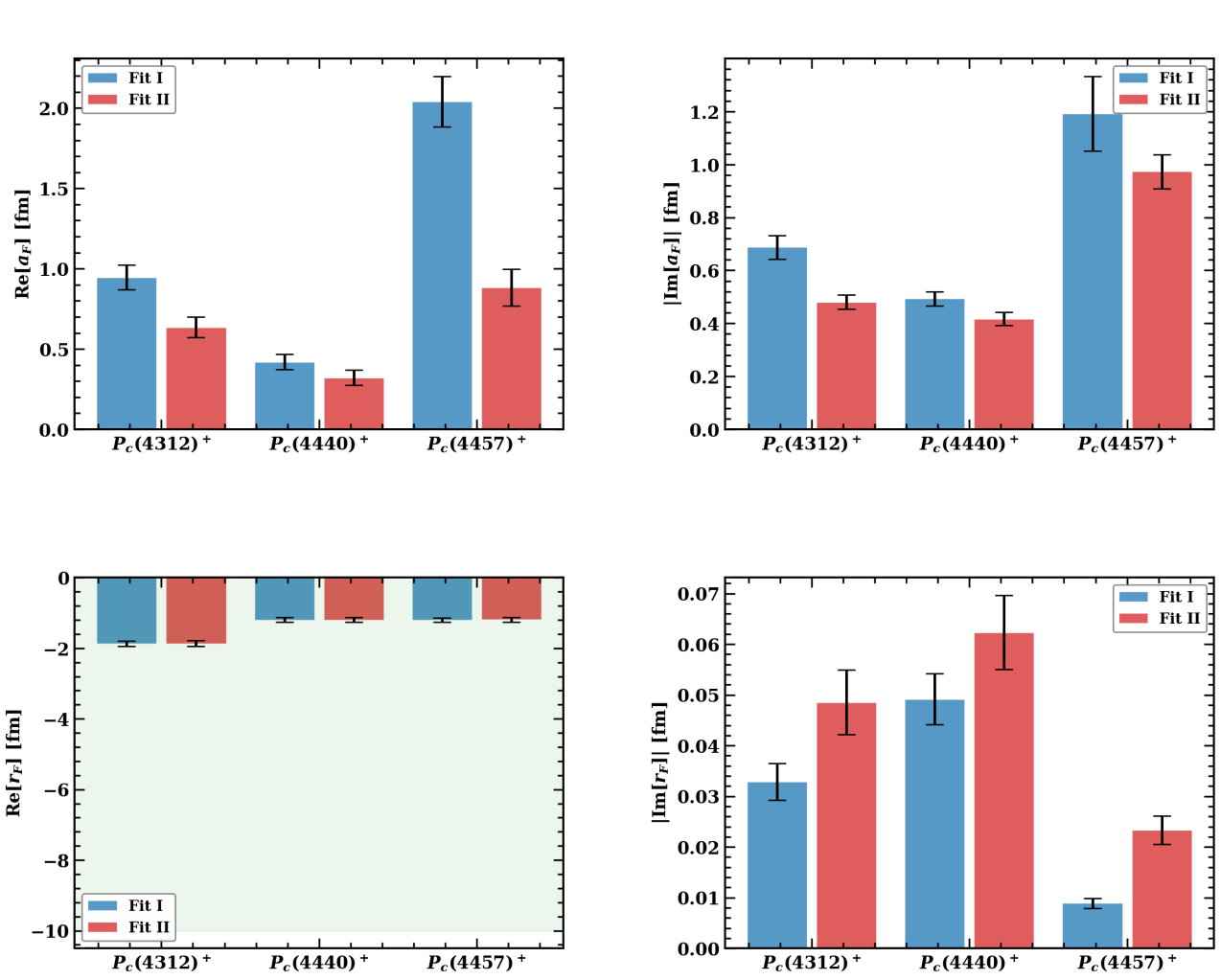

**Figure 19:** ERE parameter bar chart with bootstrap uncertainties.

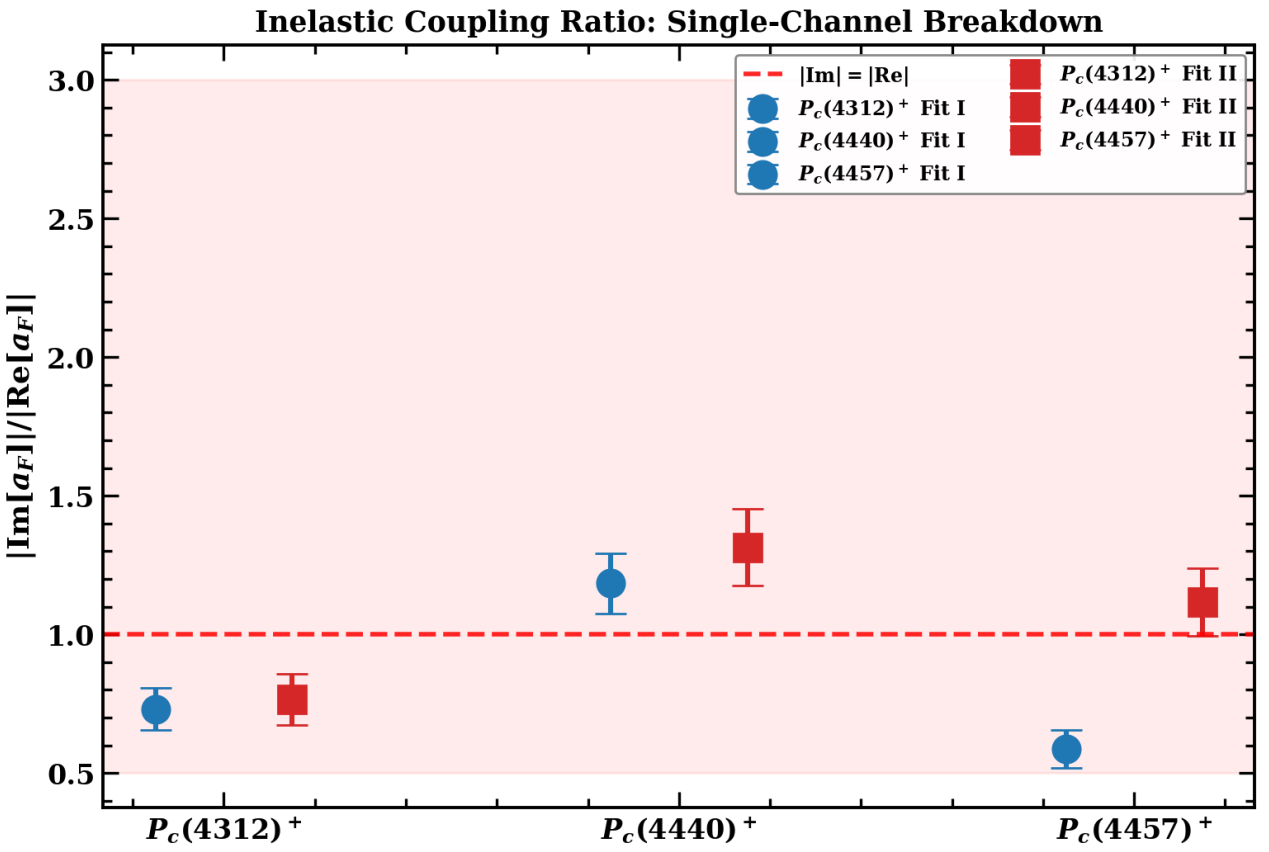

**Figure 20:** Inelastic coupling ratio $|\mathrm{Im}[a_F]|/|\mathrm{Re}[a_F]|$.

To illustrate the sensitivity of the effective-range extraction to the fitting methodology, Table 8 compares $\mathrm{Re}[r_F]$ for the $P_c(4312)^+$ (Fit I) across three distinct treatments: the isolated-signal fit with real couplings, the global fit with real couplings, and the global fit with phase-marginalized couplings.

The central values obtained from the isolated-signal and global fits are in close agreement within the real-coupling scenario, as expected given that both analyses effectively constrain the coupling phases to fixed values. For the central value of $\mathrm{Re}[r_F]$, the isolated signal and global fit approaches (real-coupling treatment) are therefore equally trustworthy; the distinctive contribution of the present paper is to make explicit, via the phase-marginalised comparison, how strongly that agreement depends on the shared real-coupling assumption.

**Table 8:** Three-way comparison of $\mathrm{Re}[r_F]$ [fm] for $P_c(4312)^+$, Fit I.

| Method | Coupling type | $\mathrm{Re}[r_F]$ [fm] |
|---|---|---|
| Isolated signal | real couplings | $-1.86 \pm 0.00$ |
| Global fit | real couplings | $-1.86 \pm 0.09$ |
| Global fit | phase-marginalised | $0.01 \pm 1.34$ |

### 6.2. Systematic uncertainties

The dominant systematic uncertainty in the ERE parameters is the choice of channel-2 threshold (Fit I vs. Fit II), which shifts $E_{\mathrm{BW}}$ by 4-6 MeV per state (Sec. 4.1) and propagates into a $\sim$ 20-30% shift in $|\mathrm{Re}[a_F]|$ between the two fit variants. Additional, currently unquantified sources of systematic uncertainty include the background polynomial order and functional form (Sec. 3.4), the width of the regularising prior of Eq. (25) (partially quantified in Table 2), the non-relativistic approximation in $T_{\mathrm{NR}}$, and the reliability of the gradient-free minimiser in the degenerate 24-parameter space (Table 4). A systematic scan over the background polynomial order, and a comparison with the $S$-wave background parametrisation used internally by the LHCb collaboration, would substantially strengthen the systematic-uncertainty budget of this analysis and are identified as concrete follow-up work.

However, several limitations should be noted. In particular, an $E_{\mathrm{BW}}$-$|g_2|$ degeneracy remains in the present analysis and is regularized rather than fully resolved through the explicit, sensitivity-tested soft prior described in Sec. 2.5. The coupling-phase degeneracy is quantified in Table 6, where phase marginalisation is found to increase $\sigma(\mathrm{Re}[r_F])$ by more than an order of magnitude and to remove the statistical significance of the molecular signature. In addition, the analysis exhibits some sensitivity to local minima, with the $\chi^2_{\mathrm{data}}/\nu$ values obtained from eight Nelder-Mead multi-starts varying by 21-43% (Table 4); a more exhaustive global optimisation using, for example, differential evolution or simulated annealing has not yet been performed. The bootstrap procedure described in Sec. 3.2 samples around the best-fit solution rather than performing a complete refit for every resampled dataset, representing a perturbative approximation adopted for computational tractability. The results also exhibit some dependence on the background model, as illustrated by the background-only fit in Fig. 2, although this dependence has not yet been quantified through a systematic scan over polynomial orders.

## 7. Conclusion and Outlook

We have presented a global two-channel Flatté amplitude analysis of the complete LHCb Run 1+2 $J/\psi\, p$ invariant-mass spectrum, simultaneously describing the $P_c(4312)^+$, $P_c(4440)^+$, and $P_c(4457)^+$ structures with fully complex couplings and a sixth-order polynomial background. From the fitted parameters, we have derived the corresponding complex effective-range-expansion parameters in closed analytic form and propagated their statistical uncertainties through non-parametric bootstrap resampling. Under the real-coupling assumption, all three states yield $\mathrm{Re}[r_F] < 0$ for both fit variants, with central values consistent with independent theoretical determinations and supportive of the $\Sigma_c\bar{D}^{(*)}$ molecular interpretation. This interpretation, however, is not robust against coupling-phase marginalisation: allowing the phases $\varphi_{1,2}$ to vary over their physical ranges removes both the definite sign and the statistical significance of $\mathrm{Re}[r_F]$, highlighting the importance of phase assumptions in extracting near-threshold scattering parameters. We also find a non-negligible overlap between the $P_c(4440)^+$ and $P_c(4457)^+$ lineshapes, with an overlap integral of $0.57$, demonstrating that the incoherent-sum approximation can become quantitatively relevant for closely spaced structures. Finally, the observed sensitivity to local minima, background modelling, and resolution treatment indicates that the systematic uncertainties associated with the global fitting strategy are not yet fully quantified.

These findings point to several methodological challenges inherent in extracting near-threshold scattering information from invariant-mass spectra. In particular, the $E_{BW}$-$|g_2|$ and coupling-phase degeneracies reflect intrinsic parameter correlations of the adopted Flatté parametrization and cannot be eliminated solely through improvements in numerical minimization; rather, they require additional experimental constraints or alternative inference strategies. The sensitivity to local minima and the computational cost of repeated non-convex fits, together with the dependence on the chosen background and resolution models, further motivate complementary data-driven approaches. A natural next step is therefore to investigate supervised machine-learning methods trained on synthetic ensembles of Flatté lineshapes spanning the relevant parameter space, including coupling phases, background shapes, and detector resolution. Such an approach could provide a rapid inference mechanism and, when properly calibrated, offer an independent assessment of the parameter degeneracies identified here. In particular, the real-coupling and phase-marginalised results obtained in this work provide a quantitative benchmark for future machine-learning analyses: a reliable model should reproduce the corresponding loss of constraining power on $r_F$ when the coupling phases are left unconstrained. This provides a well-defined direction for future work toward more flexible and computationally efficient inference of near-threshold exotic-hadron amplitudes.

## Acknowledgments

We thank the LHCb collaboration for making the dataset publicly available through HEPData. This work received funding from the Post-Doctoral Program of the Swiss Government Excellence Scholarship, Grant No. 2025.0419. Computational resources were provided by the University of Geneva.